\documentclass{acmart}
\usepackage{multirow}
\usepackage{float}
\usepackage{subfigure}
\usepackage{enumerate}
\usepackage{enumitem}
\usepackage{stfloats}
\usepackage{hyperref}
\usepackage{breakurl}
\usepackage{diagbox}
\usepackage{balance}
\usepackage{titlesec}
\usepackage{verbatimbox}
\usepackage{ulem}
\usepackage{indentfirst}
\usepackage[linesnumbered,ruled,vlined]{algorithm2e}
\usepackage{cleveref}
\usepackage{colortbl}
\usepackage{tablefootnote}

\crefname{section}{§}{§§}

\definecolor{best}{RGB}{255, 159, 69}
\definecolor{gray}{RGB}{221, 221, 221}

\AtBeginDocument{%
  \providecommand\BibTeX{{%
    \normalfont B\kern-0.5em{\scshape i\kern-0.25em b}\kern-0.8em\TeX}}}

\def\new{\color{black}}
\def\newxn{\color{black}}
\def\rev{\color{black}}
\newcommand\model{BiGeaR++}

\def\emb{\boldsymbol} 
\DeclareMathOperator\sign{sign}
\DeclareMathOperator\argmin{argmin}

\begin{document}
\title[\model: Learning Binarized Representations with Pseudo-positive Sample Learning Enhancement]{Learning Binarized Representations with Pseudo-positive Sample Enhancement for Efficient Graph Collaborative Filtering}


\author{Yankai Chen}
\affiliation{
 \institution{Cornell University}
 \country{USA}
}
\email{yankaichen@acm.org}

\author{Yue Que}
\affiliation{%
  \institution{City University of Hong Kong}
  \country{Hong Kong SAR}
 }
\email{yueque2-c@my.cityu.edu.hk}

\author{Xinni Zhang}
\affiliation{
 \institution{The Chinese University of Hong Kong}
 \country{Hong Kong SAR}
}
\email{xnzhang23@cse.cuhk.edu.hk}

\author{Chen Ma}
\affiliation{%
  \institution{City University of Hong Kong}
  \country{Hong Kong SAR}
 }
\email{chenma@cityu.edu.hk}

\author{Irwin King}
\affiliation{
 \institution{The Chinese University of Hong Kong}
 \country{Hong Kong SAR}
}
\email{king@cse.cuhk.edu.hk}

\renewcommand{\shortauthors}{Yankai Chen et al.}


\begin{abstract}
{\new 
Learning vectorized embeddings is fundamental to many recommender systems for user-item matching. 
To enable efficient online inference, \textit{representation binarization}, which embeds latent features into compact binary sequences, has recently shown significant promise in optimizing both memory usage and computational overhead. 
However, existing approaches primarily focus on \textit{numerical quantization}, neglecting the associated \textit{information loss}, which often results in noticeable performance degradation.
To address these issues, we study the problem of graph representation binarization for efficient collaborative filtering. 
Our findings indicate that explicitly mitigating information loss at various stages of embedding binarization has a significant positive impact on performance. 
Building on these insights, we propose an enhanced framework, \model, which specifically leverages supervisory signals from \textit{pseudo-positive samples}, incorporating both real item data and latent embedding samples.
Compared to its predecessor BiGeaR, \model~introduces a fine-grained inference distillation mechanism and an effective embedding sample synthesis approach. 
Empirical evaluations across five real-world datasets demonstrate that the new designs in \model~work seamlessly well with other modules, delivering substantial improvements of around 1\%$\sim$10\% over BiGeaR and thus achieving state-of-the-art performance compared to the competing methods.
}
{\rev Our implementation is available at \url{https://github.com/QueYork/BiGeaR-SS}.}
\end{abstract}




\maketitle

\section{Introduction}
Recommender systems, designed to deliver personalized information filtering~\cite{pinsage,li2024recent}, are broadly applicable across various Internet-based platforms. 
Collaborative Filtering (CF), which leverages past user-item interactions to learn vectorized user-item representations (i.e., embeddings) for predictions, continues to be a core method for effective personalized recommendation~\cite{lightgcn,ngcf}. 
To maintain the efficient inference for rapid expansion of data, \textit{representation binarization} has recently emerged as a promising solution.
Technically, it works by quantizing the latent features of users and items, transforming continuous full-precision representations into discrete binarized versions.
These binarized representations enable significant reductions in model size and speed up inference, leveraging low-bit arithmetic on devices where CPUs are generally more cost-effective than high-end GPUs~\cite{banner2018scalable,bahri2021binary}.

\begin{figure}[tp]
\centering
\begin{minipage}{1\textwidth}
\includegraphics[width=\linewidth]{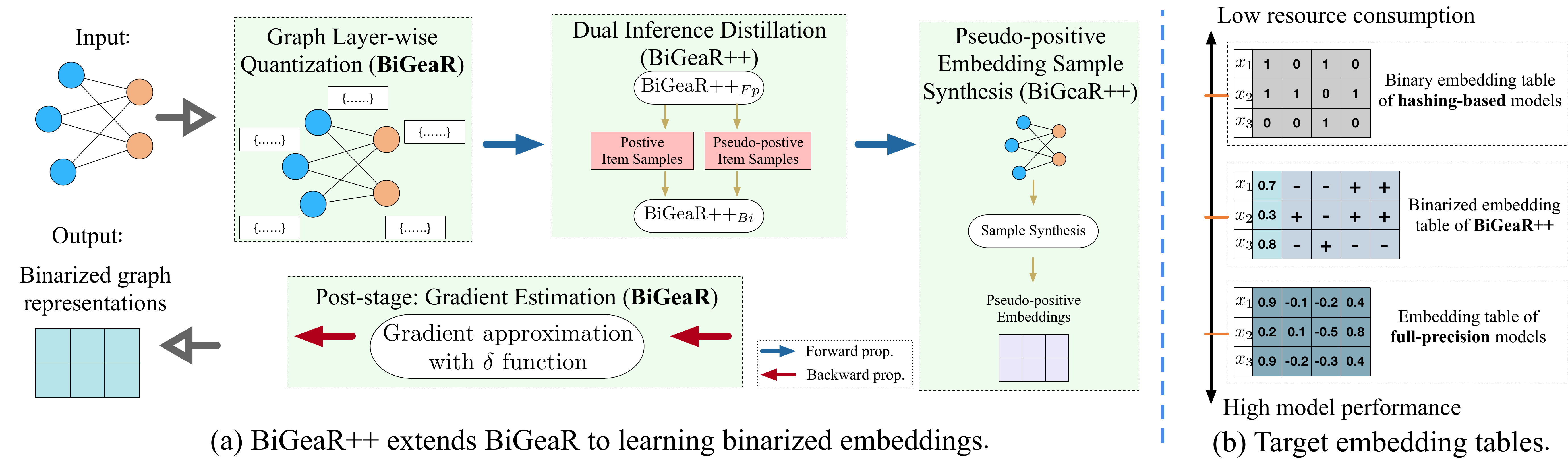}
\end{minipage} 
\caption{Illustration of \model.}
\label{fig:intro}
\end{figure} 

{\new 
Despite its promising potential, simply stacking CF methods with binarization techniques is trivial with 
large performance degradation in Top-K recommendation~\cite{hashgnn,kang2019candidate}.
To address this issue, we have progressively analyzed the root cause as stemming from the severe \textit{information loss}: i.e., \textit{limited expressivity of latent features}, \textit{degraded ranking capability}, \textit{inaccurate gradient estimation}~\cite{chen2022learning}. 
(1) Firstly, conventional methods use the $\sign(\cdot)$ function for fast embedding binarization~\cite{hashgnn,wang2017survey,lin2017towards,chen2022effective}. However, this only preserves the sign (+/-) of each embedding entry, resulting in binarized representations that are less informative compared to their full-precision counterparts. 
(2) Secondly, previous work often overlooks the discrepancy of ranking inference between full-precision and binarized embeddings~\cite{kang2019candidate,hashgnn}.
Failing to capture such information may thus lead to diminished ranking capability and sub-optimal recommendation performance.
(3) Thirdly, previous work usually adopts \textit{Straight-Through} \textit{Estimator} (STE) \cite{bengio2013estimating} to estimate the gradients for $\sign(\cdot)$, which may be inaccurate and result in inconsistent optimization directions during model training~\cite{chen2022learning}.
Therefore, we have developed a model namely \textbf{BiGeaR} to address these problems by introducing multi-faceted techniques within Graph Convolutional Network (GCN) framework~\cite{kipf2016semi,graphsage,pinsage,zhang2023mitigating}, sketching graph nodes (i.e., users and items) with binarized representations.
In general, BiGeaR facilitates nearly one bit-width representation compression and achieves state-of-the-art performance.

While our previous work successfully achieved the effectiveness-efficiency balance, it also revealed avenues for further improvement, particularly in the area of \textit{model optimization with negative data samples}. 
In the conventional learning paradigms~\cite{lightgcn,ngcf,hashgnn}, negative data samples typically refer to {\rev non-interacted} items, which are assumed to represent users' disliked preferences.
However, we contend that within these negative data samples, users' preferences may exhibit varying degrees of disinterest, which can serve as additional supervisory signals for the model.
Based on this assumption, we advance the study of graph representation binarization by introducing the enhanced model \model. 
Expanding upon its predecessor architecture, \model~places a stronger emphasis on \textit{enhancement from learning pseudo-positive samples} that are essentially negative but share the most informative supervisory signals during model learning.
In the context of embedding binarization, \model's learning of such pseudo-positive samples is twofold as follows:
\begin{itemize}[leftmargin=*]
\item In representation binarization, \model~explicitly captures the disparity between full-precision and binarized embeddings in terms of their ranking inference towards pseudo-positive \textit{real} data samples.
 
\item Furthermore, beyond the usage of real samples, \model~further introduces a pseudo-positive embedding sample synthesis approach that ultimately generates hard-to-distinguish yet highly informative latent embedding samples, which significantly enhances the model's ability to achieve more accurate predictions.
\end{itemize}
}

{\rev 
To summarize, as shown in Figure~\ref{fig:intro}(a), \model~technically consists of four designs at different stages of binarized representation learning.
Specifically, \model~directly inherits the following designs from the predecessor BiGeaR:
}

\begin{itemize}[leftmargin=*]
\item At the information aggregation stage, we introduce the \textbf{graph layer-wise quantization}, which progressively binarizes user-item features from low- to high-order interactions, capturing different semantic levels. Our analysis shows that this layer-wise quantization achieves a \textit{magnification effect} of \textit{feature uniqueness}, effectively mitigating the limited expressivity commonly associated with embedding binarization. Empirical results further demonstrate that this approach significantly enhances the informativeness of the quantized features, proving to be more effective than simply increasing embedding sizes in conventional methods~\cite{hashgnn,lin2017towards,qin2020forward,rastegari2016xnor}.

\item As for the model backpropagation stage, we propose leveraging an approximation of the \textit{Dirac delta function} (i.e., $\delta$ function)~\cite{bracewell1986fourier} for more \textbf{accurate gradient estimation}. Unlike the Straight-Through Estimator (STE), our gradient estimator ensures consistent optimization directions for $\sign(\cdot)$ during both forward and backward propagation. Empirical results demonstrate its superiority over other gradient estimation methods.
\end{itemize}

{\rev 

In addition to them, \model~further introduces the following new designs:

\begin{itemize}[leftmargin=*]

\item During the embedding quantization stage, \model~ introduces the \textbf{dual inference distillation} to develop the fine-grained ranking capability inheritance.
We firstly extract \textit{positive} training samples, i.e., interacted items, to guide the binarized embeddings in making accurate predictions to them.
To explore the additional supervisory signals, we then particularly extract \textit{pseudo-positive} \textit{training data} \textit{samples}, i.e., not interacted with but highly recommended items, based on the full-precision embeddings of \model.
These pseudo-positive samples serve as additional regularization targets to the quantized embeddings, allowing the full-precision embeddings to distill ranking knowledge to the binarized ones and achieve similar inference outcomes.

\item In the model training stage, we further leverage the supervisory signals from pseudo-positive samples.
Different from the previous step that uses real items mainly for inference distillation, we propose an effective \textbf{pseudo-positive sample synthesis} approach that operates in the latent embedding space.
These synthesized pseudo-positive samples are not actual items but latent embeddings with the most informative and difficult-to-distinguish features.
We implement the sample synthesis in both the full-precision and binarized spaces to ensure comprehensive training of \model.
Our empirical analysis demonstrates the superiority of this approach compared to other sample synthesis methods~\cite{MixGCF2021,DINS2023} across various datasets.
\end{itemize}
}

\paragraph{\textbf{Experimental Results.}}
Extensive experiments conducted on five real-world benchmarks demonstrate that {\new our \model~delivers significant performance improvements than its predecessor BiGeaR with around 1.08\%$\sim$10.26\% improvement.
Specifically, it continues to outperform the state-of-the-art binarized recommender models by 32.47\%$\sim$54.49\% and 25.46\%$\sim$50.86\% in terms of Recall and NDCG, respectively. Moreover, \model~achieves 98.11\%$\sim$106.61\% and 98.13\%$\sim$108.26\% of the recommendation capability when compared to the best-performing full-precision models}.
By decomposing the prediction formula into bitwise operations, \model~substantially reduces the number of floating-point operations (FLOPs), providing a theoretically grounded acceleration for online inference. As a result, \model~achieves over 8$\times$ faster inference speed and space compression compared to its full-precision counterparts.

\paragraph{\textbf{Discussion and Organization.}}
It is worth noting that \model~is conceptually related to \textit{hashing-based} models (i.e., learning to hash)~\cite{kang2021learning,kang2019candidate}, as binary hashing can be seen as a form of 1-bit quantization. However, as illustrated in Figure~\ref{fig:intro}(b), the two approaches are driven by different motivations. Hashing-based models are typically designed for efficient candidate generation, followed by full-precision \textit{re-ranking} algorithms to ensure accurate predictions. In contrast, \model~operates in an \textit{end-to-end} manner, with the goal of making predictions entirely within the proposed architecture. Thus, while \model~is \textit{technically} related to hashing-based models, it is \textit{motivationally} distinct. We present \model~ methodology and model analysis in~\cref{sec:end} and~\cref{sec:analysis}. The experimental results and related work are discussed in~\cref{sec:exp} and~\cref{sec:work}, followed by the conclusion in~\cref{sec:con}.

\section{\rev \model~Methodology}
\label{sec:end}
\begin{figure*}[tp]
\begin{minipage}{1\textwidth}
\includegraphics[width=\linewidth]{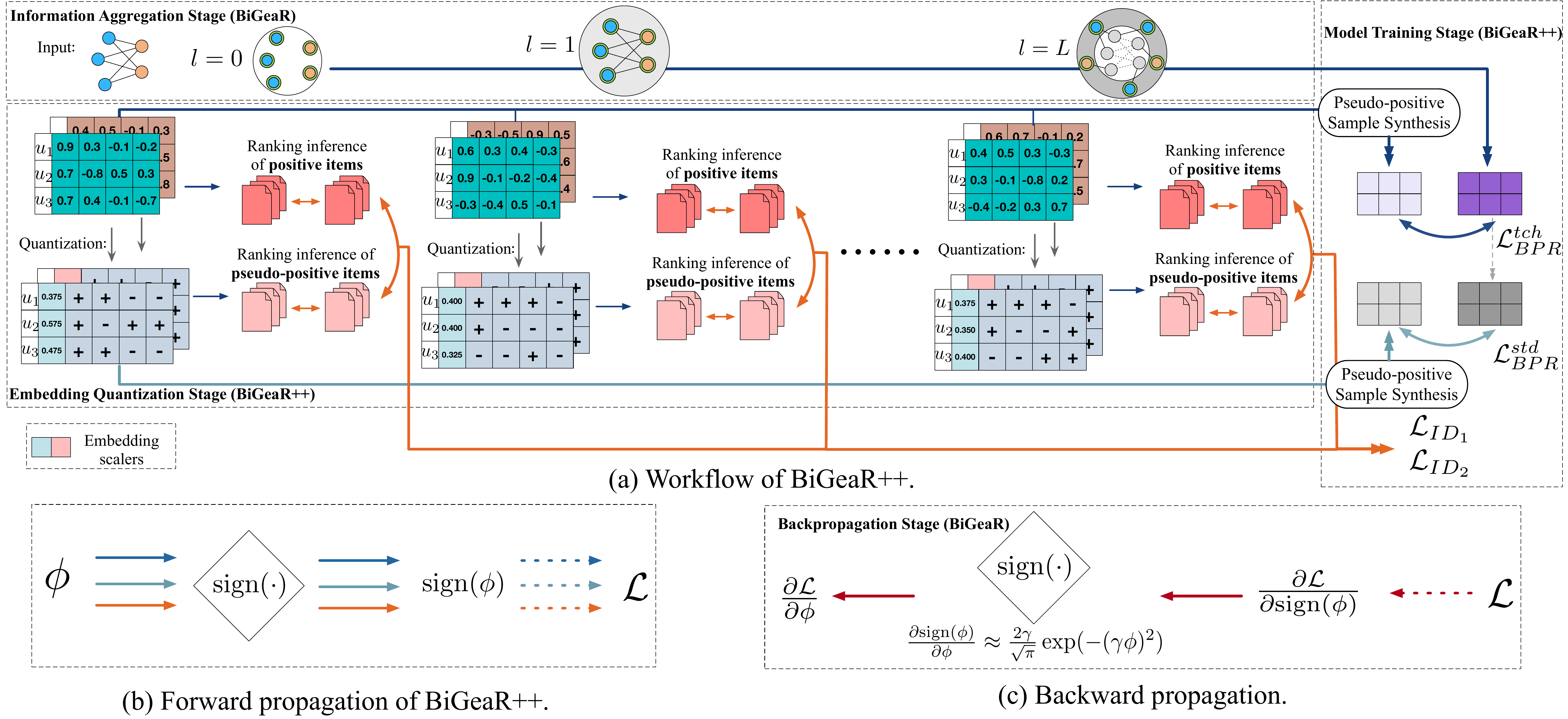}
\end{minipage} 
\caption{\new \model~first pre-trains the full-precision embeddings and then triggers the (1) graph layer-wise quantization, (2) dual inference distillation, (3) pseudo-positive sample synthesis, and (4) accurate gradient estimation to learn the binarized representations (Best view in color).}
\label{fig:framework}
\end{figure*}

In this section, we formally introduce:
\textit{(1) graph layer-wise quantization for feature magnification;
{\new (2) dual inference distillation for ranking capability inheritance;
(3) pseudo-positive sample synthesis for informative ranking learning;
}
(4) gradient estimation for better model optimization.}
\model~framework is illustrated in Figure~\ref{fig:framework}(a).
{\new We explain all key notations in Table~\ref{tab:notation}.}

\begin{table}[th]
\caption{\new Notations and meanings. }
\label{tab:notation}
{\new 
\begin{tabular}{ll}
  \hline
      {\bf Notation} & {\bf Explanation}\\
  \hline
      $d$, $L$    & {Embedding dimensions and graph convolution layers.} \\
      $\mathcal{U}$, $\mathcal{I}$ & Collection of users and items. \\
      $\mathcal{N}(x)$ & {Neighbors of node $x$.} \\
      $\boldsymbol{v}_{x}^{(l)}$ & {Full-precision embedding of node $x$ at $l$-th convolution.}\\
      $\boldsymbol{q}_{x}^{(l)}$ & {Binarized embedding of node $x$ at $l$-th quantization.}\\
      $\alpha_{x}^{(l)}$ & {$l$-th embedding scaler of node $x$.}\\
      $\mathcal{A}_x$ and $\mathcal{Q}_x$ & {Binarized embedding table of $x$ learned by \model. } \\
      $w_l$   & {$l$-th weight in predicting matching score.} \\
      $y_{u,i}$     & {A scalar indicates the existence of user-item interaction.} \\
      $\widehat{y}^{tch}_{u,i}$  &   {Predicted score based on full-precision embeddings.}\\
      $\widehat{y}^{std}_{u,i}$  &   {Predicted score based on binarized embeddings.}\\
      $\emb{\widehat{y}}^{tch,\,(l)}_{u}$  & {Predicted scores of $u$ based on $l$-th embeddings segments.}\\
      $\emb{\widehat{y}}^{std,\,(l)}_{u}$  & {Predicted scores of $u$ based on $l$-th quantized segments.}\\
      $S_{tch}^{(l)}(u)$ & {Pseudo-positive training samples of $u$.}\\
      $c$ & Uniform distribution scaler. \\
      $w_k$   &   {$k$-th weight in inference distillation loss.}\\
      $\mathcal{L}_{BPR}^{tch}$, $\mathcal{L}_{BPR}^{std}$  & {BPR loss based on full-precision and binarized scores.} \\
      $\mathcal{L}_{ID_1}$, $\mathcal{L}_{ID_2}$ & {Dual inference distillation loss terms.} \\
      $\mathcal{E}_{can}$, $\mathcal{E}_{neg}$ & Candidate embedding pool and negative embedding set. \\ 
      $\boldsymbol{e}_i^{(l), -}$, $\boldsymbol{e}_i^{-}$ & Synthesized pseudo-positive embedding at the $l$-th layer and the final one. \\
      $\mathcal{L}$ & {Objective function of \model.} \\
      $u(\cdot)$, $\delta(\cdot)$ & Unit-step function and Dirac delta function.\\
      $\lambda$, $\lambda_1$, $\lambda_2$, $\gamma$, $\eta$ & {Hyper-parameters and learning rate.} \\
\hline
\end{tabular}
}
\end{table}

\textbf{Preliminaries: graph convolution.} 
The general approach is to learn node representations by iteratively propagating and aggregating latent features from neighbors through the graph topology~\cite{wu2019simplifying,zhang2025topology,kipf2016semi,wang2025bangs}. We adopt the convolution paradigm operating in the continuous space from LightGCN~\cite{lightgcn}, which has recently demonstrated strong performance. Let {$\boldsymbol{v}^{(l)}_u$} and {$\boldsymbol{v}^{(l)}_i \in \mathbb{R}^{d}$} denote the continuous feature embeddings of user $u$ and item $i$ after $l$ layers of information propagation, and let {$\mathcal{N}(x)$} represent the neighbor set of node $x$. These embeddings are iteratively updated using information from the ($l$-$1$)-th layer as follows:
\begin{equation}
\label{eq:gcn}
\boldsymbol{v}^{(l)}_u = \sum_{i\in \mathcal{N}(u)} \frac{1}{\sqrt{|\mathcal{N}(u)|\cdot|\mathcal{N}(i)|}}\boldsymbol{v}^{(l-1)}_i, \ \ \boldsymbol{v}^{(l)}_i = \sum_{u\in \mathcal{N}(i)} \frac{1}{\sqrt{|\mathcal{N}(i)|\cdot|\mathcal{N}(u)|}}\boldsymbol{v}^{(l-1)}_u.
\end{equation}

\subsection{\textbf{Graph Layer-wise Quantization}}
We propose the \textit{graph layer-wise quantization} by computing both \textbf{quantized embeddings} and \textbf{embedding scalers}: (1) The quantized embeddings approximate the full-precision embeddings using $d$-dimensional binarized codes (i.e., $\{-1, 1\}^d$); and (2) Each embedding scaler captures the value range of the original embedding entries. Specifically, during graph convolution at each layer, we track the intermediate information (e.g., {$\boldsymbol{v}^{(l)}_u$}) and perform layer-wise 1-bit quantization in parallel, as follows:
\begin{equation}
\boldsymbol{q}_u^{(l)} = \sign\big(\boldsymbol{v}^{(l)}_u\big), \ \ \boldsymbol{q}_i^{(l)} = \sign\big(\boldsymbol{v}^{(l)}_i\big),
\end{equation}
where embedding segments {$\boldsymbol{q}_u^{(l)}$, $\boldsymbol{q}_i^{(l)}$ $\in$ $\{-1, 1\}^d$} retain the node latent features directly from {$\boldsymbol{v}^{(l)}_u$} and {$\boldsymbol{v}^{(l)}_i$}.
To equip with the layer-wise quantized embeddings, we introduce a layer-wise positive embedding scaler for each node (e.g., {$\alpha_u^{(l)}$ $\in$ $\mathbb{R}^+$}), such that {$\boldsymbol{v}^{(l)}_u$ $\doteq$ $\alpha_u^{(l)}$$\boldsymbol{q}^{(l)}_u$}. For each entry in {$\alpha_u^{(l)} \boldsymbol{q}^{(l)}_u$}, the values are binarized as either {$-\alpha_u^{(l)}$} or {$\alpha_u^{(l)}$}.
In this work, we compute the $\alpha_u^{(l)}$ and $\alpha_i^{(l)}$ as the L1-norms of $\boldsymbol{v}^{(l)}_u$ and $\boldsymbol{v}^{(l)}_i$, respectively, as follows:
\begin{equation}
\alpha_u^{(l)} = \frac{1}{d}\cdot||\boldsymbol{v}_u^{(l)}||_1, \ \ \alpha_i^{(l)} = \frac{1}{d} \cdot ||\boldsymbol{v}_i^{(l)}||_1.
\end{equation}
Instead of making {$\alpha_u^{(l)}$} and {$\alpha_i^{(l)}$} learnable, this \textit{deterministic} computation approach is simple yet effective in providing the scaling functionality while significantly reducing the parameter search space. After $L$ layers of quantization and scaling, we construct the following \textbf{binarized embedding table} for each graph node $x$ as:
\begin{equation}
\mathcal{A}_x = \{\alpha_x^{(0)}, \alpha_x^{(1)}, \cdots, \alpha_x^{(L)}\}, \ \ \mathcal{Q}_x = \{\boldsymbol{q}^{(0)}_x, \boldsymbol{q}^{(1)}_x, \cdots, \boldsymbol{q}^{(L)}_x\}.
\end{equation}
From a technical perspective, \model~binarizes intermediate semantics at different layers of the \textit{receptive field}~\cite{velivckovic2017graph,wu2020comprehensive} for each node. This process effectively achieves a \textbf{magnification effect of feature uniqueness}, enhancing user-item representations through exploration of the interaction graph. Further analysis of this effect is provided in~\cref{sec:necessity}.

\subsection{Prediction Acceleration}
\textbf{Model Prediction.}
Based on the learned embedding table, we predict the matching scores by adopting the inner product:
\begin{equation}
\label{eq:score}
\widehat{y}_{u,i} =  \big<f(\mathcal{A}_u, \mathcal{Q}_u), f(\mathcal{A}_i, \mathcal{Q}_i)\big>,
\end{equation}
where function $f(\cdot, \cdot)$ in this work is implemented as:  
\begin{equation}
\label{eq:useQ}
f(\mathcal{A}_u, \mathcal{Q}_u) = \Big|\Big|_{l=0}^L w_l  \alpha_u^{(l)} \boldsymbol{q}^{(l)}_u, \ \ f(\mathcal{A}_i, \mathcal{Q}_i) = \Big|\Big|_{l=0}^L w_l \alpha_i^{(l)} \boldsymbol{q}^{(l)}_i. 
\end{equation}
$\big|\big|$ denotes the concatenation of binarized embedding segments, with weight $w_l$ measuring the contribution of each segment to the prediction. The weight $w_l$ can either be defined as a hyper-parameter or treated as a learnable variable (e.g., using an attention mechanism~\cite{velivckovic2017graph}). In this work, we set $w_l$ $\propto l$, meaning that $w_l$ increases linearly from lower layers to higher layers, primarily for computational simplicity and stability.

\begin{figure}[tp]
\begin{minipage}{1\textwidth}
\includegraphics[width=\linewidth]{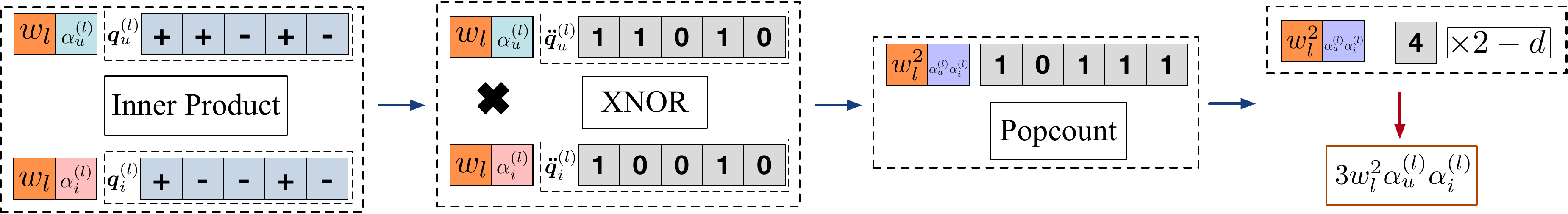}
\end{minipage} 
\caption{\new Acceleration with bitwise operations.}
\label{fig:bitwise}
\end{figure}

\textbf{Computation Acceleration.}
Notice that for each segment of { $f(\mathcal{A}_u, \mathcal{Q}_u)$}, e.g., { $w_l\alpha_u^{(l)} \boldsymbol{q}^{(l)}_u$}, entries are binarized by two values (i.e., {$-w_l  \alpha_u^{(l)}$} or {$w_l  \alpha_u^{(l)}$}). 
Thus, we can achieve the prediction acceleration by decomposing Equation~(\ref{eq:score}) with \textit{bitwise operations}. 
Concretely, in practice, {$\boldsymbol{q}^{(l)}_u$} and {$\boldsymbol{q}^{(l)}_i$} will be firstly encoded into basic $d$-bits binary codes, denoted by {$\boldsymbol{{\ddot q}}^{(l)}_u, \boldsymbol{{\ddot q}}^{(l)}_i \in \{0,1\}^d$}.
Then we replace Equation~(\ref{eq:score}) by introducing the following formula:
\begin{equation}
\label{eq:bit}
\widehat{y}_{u,i} = \sum_{l=0}^{L} w_l^2\alpha_u^{(l)}\alpha_i^{(l)}\cdot \big(2{\tt Popcount} \big({\tt XNOR}(\boldsymbol{{\ddot q}}^{(l)}_u, \boldsymbol{{\ddot q}}^{(l)}_i  )\big) - d\big).
\end{equation}
Compared to the original computation method in Equation~(\ref{eq:score}), our bitwise-operation-supported prediction in Equation~(\ref{eq:bit}) significantly reduces the number of floating-point operations (\#FLOP) by utilizing {\tt Popcount} and {\tt XNOR} operations.
{\new We provide an example to illustrate the computation process in Figure~\ref{fig:bitwise}.}

\subsection{\new Dual Inference Distillation for Ranking Capability Inheritance}
\subsubsection{\textbf{Ranking Capability Inheritance.}}
To address the issue of \textit{asymmetric inference capability} between full-precision and binarized representations, we introduce \textit{self-supervised inference distillation}, allowing binarized embeddings to effectively inherit inference knowledge from their full-precision counterparts. Specifically, we treat the full-precision intermediate embeddings (e.g., {$\boldsymbol{v}_u^{(l)}$}) as the \textbf{teacher} embeddings and the quantized segments as the \textbf{student} embeddings. Given both teacher and student embeddings, we can compute their respective prediction scores, {$\widehat{y}_{u,i}^{tch}$} and {$\widehat{y}_{u,i}^{std}$}. For Top-K recommendation, our goal is to minimize the discrepancy between them, formulated as:
\begin{equation}
\argmin \mathcal{D}(\widehat{{y}}_{u,i}^{tch}, \widehat{{y}}_{u,i}^{std}).
\end{equation}

{\new
\subsubsection{\textbf{Dual Inference Distillation.}} 
As pointed by early work~\cite{chen2022learning}, a straightforward implementation of $\mathcal{D}$ from the conventional \textit{knowledge distillation}~\cite{hinton2015distilling,anil2018large}.
}
A common approach is to minimize the Kullback-Leibler divergence (KLD) or mean squared error (MSE) between the teacher and student embeddings. While these methods are effective in classification tasks (e.g., visual recognition~\cite{anil2018large,xie2020self}), they may not be well-suited for Top-K recommendation as:

\begin{itemize}[leftmargin=*] 
\item First, both KLD and MSE in $\mathcal{D}$ encourage the student logits (e.g., {$\widehat{y}_{u,i}^{std}$}) to be similarly distributed to the teacher logits from a macro perspective. However, for ranking tasks, these methods may not effectively capture the relative order of user preferences toward items, which is critical for enhancing Top-K recommendation performance.

\item Second, KLD and MSE apply distillation across the entire item corpus, which can be computationally expensive for realistic model training. Given that item popularity typically follows a Long-tail distribution~\cite{park2008long,tang2018ranking}, learning the relative order of frequently interacted items at the top of ranking lists is more impactful for improving Top-K recommendation performance. 
\end{itemize}

{\new 
To effectively transfer the ranking capability for Top-K recommendation, we encourage the binarized representations to produce similar inference outputs as their full-precision counterparts. Specifically, we extract \textit{interacted training samples} along with additional \textit{pseudo-positive training samples} from the teacher embeddings to regularize the target embeddings at each convolutional layer. Pseudo-positive samples refer to items that a user has not interacted with but are more likely to be preferred compared to other {\rev non-interacted} items. In this work, we propose a dual inference distillation approach, which trains the binarized embeddings to generate similar ranking lists not only for interacted samples but also for pseudo-positive item samples.
}

{\new 
We first leverage the \textit{teacher embeddings}, i.e., full-precision ones, to calculate the teacher score { $\widehat{y}^{\,tch}_{u,i}$} with the inner product operation as follows:
\begin{equation}
\widehat{y}^{\,tch}_{u,i} = \big<\big|\big|_{l=0}^L w_l\boldsymbol{v}_u^{(l)}, \big|\big|_{l=0}^L w_l\boldsymbol{v}_i^{(l)}\big>, \text{ where } i \in \mathcal{N}(u).
\end{equation}
For each user $u$, we retrieve the layer-wise teacher inference towards all interacted items $\mathcal{N}(u)$:
\begin{equation}
\label{eq:tch_emb}
\widehat{\emb{y}}^{\,tch, (l)}_{u} = \big< w_l{\emb{v}}_u^{(l)},  w_l{\emb{v}}_i^{(l)}\big>, \text{ where } i \in \mathcal{N}(u).
\end{equation}
Based on the segment scores { $\boldsymbol{\widehat{y}}_u^{\,tch, (l)}$} at the $l$-th layer, we first rank these items based on their estimated matching scores. 
We propose the first part of our inference distillation designs as follows:
\begin{equation}
\label{eq:tch_layer_wise}
\mathcal{L}_{ID_1}(u) = \sum_{l=0}^L \mathcal{L}_{ID_1}^{(l)}\big(\widehat{\boldsymbol{y}}_u^{\,std, (l)}, \mathcal{N}(u)\big) = -\frac{1}{|\mathcal{N}(u)|} \sum_{l=0}^L \sum_{k=1}^{|\mathcal{N}(u)|} w_k \cdot \ln\sigma(\widehat{y}_{u,\mathcal{N}(u, k)}^{\,std, (l)}),
\end{equation}
The student scores {$\widehat{\emb{y}}^{,std, (l)}_{u}$} are computed based on the binarized embeddings, similar to Equation~(\ref{eq:tch_emb}). Here, {$\mathcal{N}(u, k)$} returns the $k$-th highest-scoring item from the interacted training items $\mathcal{N}(u)$. The ranking-aware weight, $w_k$, is used to dynamically adjust the importance of different ranking positions. To achieve this, $w_k$ can be modeled using a parameterized geometric distribution, which approximates the tailed item popularity~\cite{rendle2014improving}:
\begin{equation}
\label{eq:wk}
w_k = \lambda_1 \exp(-\lambda_2\cdot k),
\end{equation}
where { $\lambda_1$} and { $\lambda_2$} control the loss contribution level and sharpness of the distribution.

Additionally, for the teacher full-precision embeddings, we {\rev extract $R$ pseudo-positive training samples from real non-interacted items} for inference distillation ({\rev Hyper-parameter settings are reported in Table~\ref{tab:hyperparameter} of \cref{sec:hyper_setting}}). 
Specifically, we select the Top-$R$ items with the highest matching scores, denoted by {$S^{(l)}_{tch}(u)$}. Note that $S^{(l)}_{tch}(u)$ is a subset of items that are not part of the user's interacted items, i.e., $S^{(l)}_{tch}(u)$ $\subseteq$ $\mathcal{I}\setminus \mathcal{N}(u)$. 
The second part of the inference distillation is formulated as follows:
\begin{equation}
\mathcal{L}_{ID_2}(u) = \sum_{l=0}^L \mathcal{L}_{ID_2}^{(l)}\big(\widehat{\boldsymbol{y}}_u^{\,std, (l)}, S^{(l)}_{tch}(u)\big) = -\frac{1}{R} \sum_{l=0}^L \sum^{R}_{k=1} w_k \cdot \ln\sigma(\widehat{y}_{u,S^{(l)}_{tch}(u, k)}^{\,std, (l)}).
\end{equation}
$w_k$ is the ranking-aware weight we introduced earlier; since the samples in {$S^{(l)}_{tch}(u)$} are not necessarily all ground-truth positives, $w_k$ is used to balance their contribution to the overall loss.

Intuitively, {$\mathcal{L}_{ID_1}$} encourages the ground-truth interacted items from the full-precision embeddings to appear more frequently in the student's inference list. To strengthen this effect, {$\mathcal{L}_{ID_2}$} directly distills the teacher's knowledge of highly recommended unknown items, offering comprehensive supervision for the student embeddings to learn about all potential positive interactions. This distillation approach regularizes embedding quantization in a layer-wise manner, effectively reducing inference discrepancies and enhancing the student's recommendation accuracy.
}

{\new
\subsection{Pseudo-positive Sample Synthesis for Informative Ranking Learning}
In graph-based collaborative filtering, the conventional learning approach focuses on training the recommender model to accurately differentiate between positive (i.e., interacted) items and negative (i.e., {\rev non-interacted}) items. This distinction is crucial for improving recommendation accuracy and ensuring that the model ranks relevant items correctly for users. For instance, a widely adopted learning objective, the \textit{Bayesian Personalized Ranking} (BPR) loss~\cite{rendle2012bpr}, specifically optimizes ranking by encouraging the model to prioritize interacted items over those that have not been interacted with. We generalize its formulation as follows for illustration:
\begin{equation}
\mathcal{L}_{BPR}=-\sum_{u \in \mathcal{U}} \sum_{\substack{i \in \mathcal{N}(u) \\ j \notin \mathcal{N}(u)}} \ln \sigma\left(\widehat{y}_{u, i}-\widehat{y}_{u, j}\right).
\end{equation}
Here, $\mathcal{N}(u)$ represents the set of items that user $u$ has interacted with. However, in the standard setting, negative samples are typically real items randomly selected from all {\rev non-interacted} items, following a uniform distribution. While this method provides a simple mechanism for sampling negatives, it is often suboptimal because it doesn't prioritize selecting the most informative or challenging samples~\cite{MixGCF2021,DINS2023,ANS2023,wang2022leveraging}. This uniform sampling may lead to less effective training, as the model might not encounter harder negatives, which are crucial for refining its ability to distinguish between positive and negative items. To overcome this limitation, especially within the discrete quantization space, we propose a novel and effective pseudo-positive sample synthesis approach, as illustrated in Figure~\ref{fig:negative}, described below.

\begin{figure}[t]
\begin{minipage}{1\textwidth}
\includegraphics[width=\linewidth]{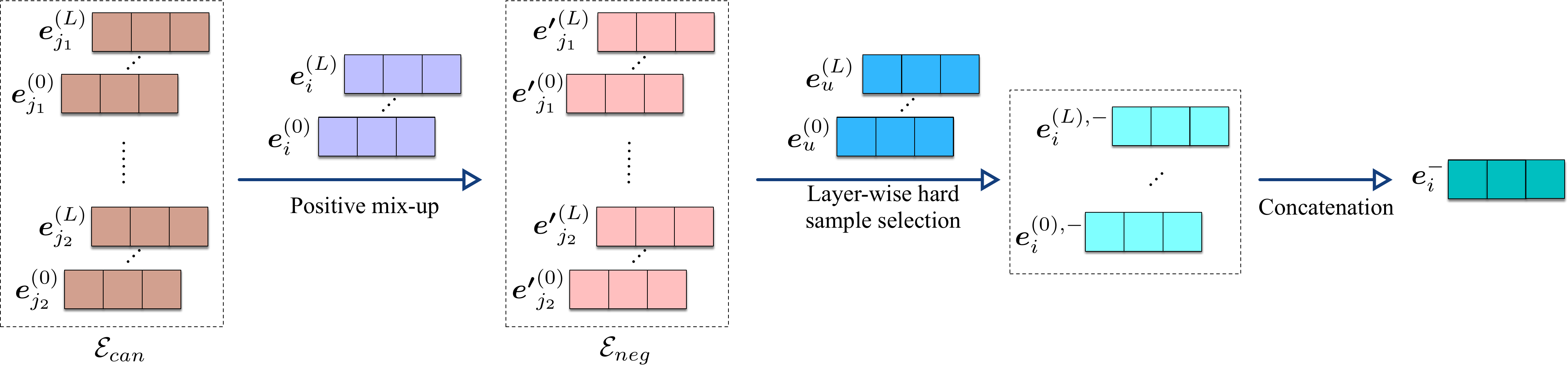}
\end{minipage} 
\caption{\new Illustration of our pseudo-positive sample synthesis approach.}
\label{fig:negative}
\end{figure}  

\subsubsection{\textbf{Pseudo-positive Sample Synthesis.}}
Instead of relying on real items for negative sampling~\cite{UltraWiki,li2024mesed}, and inspired by~\cite{MixGCF2021,DINS2023}, we develop a pseudo-positive sample synthesizer to generate highly informative and synthetic (i.e., fake) negative samples within the latent embedding space, by leveraging the existing item embeddings. This approach addresses the limitations of traditional negative sampling methods, which may struggle to effectively capture the complex boundary between positive (interacted) and negative ({\rev non-interacted}) items. The synthesizer operates in two stages: \textit{positive mix-up} and \textit{layer-wise hard sample selection}. Notably, our synthesizer functions in both full-precision and binarized spaces. For clarity in this section, we temporarily use the notation $\boldsymbol{e}$ to represent both full-precision and binarized embeddings.
\begin{itemize}[leftmargin=*]
\item In \textit{positive mix-up}, given a positive item $i$, we first sample $J$ negative items randomly, i.e., $j$ $\notin$ $\mathcal{N}(u)$. 
Their corresponding embeddings from each layer form the candidate embedding pool $\mathcal{E}_{can} = \{ \boldsymbol{e}_{j}^{(l)}|j \notin \mathcal{N}(u) \text{ and } l=0, 1, \dots, L\}$.
The size of $\mathcal{E}_{can}$ is $J \times (\textit{L+1})$. 
Then for each element $\boldsymbol{e}_{j}^{(l)} \in \mathcal{E}_{can}$, we process the \textit{mix-up} step as follows:
\begin{equation}
\boldsymbol{e'}_{j}^{(l)} = \boldsymbol{\beta}^{(l)}\circ\boldsymbol{e}_{i}^{(l)} + (\boldsymbol{1} - \boldsymbol{\beta}^{(l)})\circ\boldsymbol{e}_{j}^{(l)} ,
\end{equation} 
where $\boldsymbol{\beta}^{(l)} = [{\beta}_{0}^{(l)}, {\beta}_{1}^{(l)}, \ldots, {\beta}_{d-1}^{(l)}]$ is the dimension-wise weight vector used to mix positive information into original negative embeddings.
The ``$\circ$'' operator denotes element-wise vector multiplication. Different negative sampling methods implement $\boldsymbol{\beta}^{(l)}$ in various ways. For example, MixGCF~\cite{MixGCF2021} applies layer-wise random weights, while DINS~\cite{DINS2023} uses dimension-wise exponential ranking weights based on factor hardness. In our work, we propose constructing $\boldsymbol{\beta}^{(l)}$ as \textit{uniformly random weights across both dimensions and layers}, scaled by a control factor $c$, which limits the extent of positive signal mixing. This design helps maintain model performance stability across datasets with different sparsity levels. The formal definition is as follows:
\begin{equation}
{\beta}_{x}^{(l)} \sim \mathcal{U}(0, c) \quad \text{for} \ x = 0, 1, \ldots, (d-1) \ \ \text{and} \ \ l = 0, 1, \ldots, L.
\end{equation}

\item In \textit{layer-wise hard sample selection}, we focus on identifying the hardest samples that complicate the model's decision-making process. This approach improves the model's ability to differentiate between positive items and the most difficult-to-distinguish negative items. Notably, a recent study~\cite{yang2020understanding} theoretically shows that an effective negative sampling strategy is to select negatives based on the estimated ``positive distribution'', which reflects the user $u$'s preference for interacted items $i$. Focusing on negatives that closely resemble this positive distribution allows the model to refine its ability to distinguish between them.
To approximate this distribution, we use the inner product score to select the candidate sample with the highest score, commonly known as the hard negative selection strategy~\cite{rendle2014improving,zhang2013optimizing}. Specifically, let $\mathcal{E}_{neg}$ denote the set of embeddings generated in the previous step, i.e., $\mathcal{E}_{neg} = \{ \boldsymbol{e'}_{j}^{(l)} | j \notin \mathcal{N}(u), l = 0, 1, \dots, L \}$. We then compare the negative embeddings in $\mathcal{E}_{neg}$ to identify the hardest negative sample, denoted as $\boldsymbol{e}_{i}^{-}$. This is done by filtering out the negative embedding $\boldsymbol{e}_{i}^{(l), {-}}$ based on the inner-product-based estimated distribution at layer $l$.
\begin{equation}
\label{eq:layernega}
\boldsymbol{e}_{i}^{(l), {-}} = \mathop{\arg\max}\limits_{\boldsymbol{e'}_{j}^{(l)} \in \mathcal{E}_{neg}} \boldsymbol{e'}_{j}^{(l)} \cdot \boldsymbol{e}_{u}^{(l)},
\end{equation}
which contributes to the synthesis of our target pseudo-positive sample $\boldsymbol{e}_{i}^{-}$ as:
\begin{equation}
\label{eq:nega}
\boldsymbol{e}_{i}^{-} = \Big|\Big|_{l=0}^L \boldsymbol{e}_{i}^{(l), {-}}.
\end{equation}
\end{itemize}
The inner product score in Equation~(\ref{eq:layernega}), which measures the similarity between the user and negative samples, is computed for each candidate negative sample. The pseudo-positive sample, having the highest inner product score, is the most challenging for the model to distinguish from a positive item. These two steps extract unique information from each of the hard negatives identified, and then combine them with a high degree of diversity. This process ultimately creates synthetic yet highly informative embedding samples, enriching the model's decision-making capabilities.

\subsubsection{\textbf{Dual Space Implementation.}}
The aforementioned synthesizer could be encapsulated into the function: 
\begin{equation}
\boldsymbol{e}_{i}^{-} = \text{synthesizer}(\boldsymbol{e}_{u}^{(l)}, \boldsymbol{e}_{i}^{(l)}, \{\boldsymbol{e}_{j}^{(l)}\})^{l=0, \cdots, L}_{i \in \mathcal{N}(u), j \notin \mathcal{N}(u)} .
\end{equation}
After our graph layer-wise convolution and quantization, we operate $\text{synthesizer}(\cdot)$ jointly for both full-precision $\boldsymbol{v}_{u}^{(l)}$ and binarized $\boldsymbol{q}_{u}^{(l)}$ as:
\begin{equation}
\boldsymbol{v}_{i}^- = \text{synthesizer}(\boldsymbol{v}_{u}^{(l)}, \boldsymbol{v}_{i}^{(l)}, \{\boldsymbol{v}_{j}^{(l)}\}^{l=0, \cdots, L}_{i \in \mathcal{N}(u), j \notin \mathcal{N}(u)}, \quad
\boldsymbol{q}_{i}^- = \text{synthesizer}(\boldsymbol{q}_{u}^{(l)}, \boldsymbol{q}_{i}^{(l)}, \{\boldsymbol{q}_{j}^{(l)}\}^{l=0, \cdots, L}_{i \in \mathcal{N}(u), j \notin \mathcal{N}(u)} .
\end{equation}
Based on these pseudo-positive samples, we train the full-precision and binarized embeddings separately. 
Specifically, let $\sigma$ denote the activation function (e.g., Sigmoid). For the full-precision teacher embeddings, we minimize the BPR loss between the ground-truth items and the constructed negative samples, as follows:
\begin{equation}
\label{eq:hd-bpr}
\mathcal{L}^{tch}_{BPR} = -\sum_{u \in \mathcal{U}} \sum_{i\in \mathcal{N}(u)} \ln \sigma(\widehat{y}^{\,tch}_{u,i} - \widehat{y}^{\,tch, -}_{u,i}), \text{ where  } \widehat{y}^{\,tch, -}_{u,i} = \big<\big|\big|_{l=0}^L w_l\boldsymbol{v}_u^{(l)}, \boldsymbol{v}_i^{-}\big>.
\end{equation}
where we use $\widehat{y}^{\,tch, -}_{u,i}$ to refer to the score between user embedding and synthesized embeddings that differentiates $\widehat{y}^{\,tch}_{u,i}$.
Please notice that we only disable binarization and its associated gradient estimation (introduced later in Section~\cref{sec:gradient}) in training full-precision embeddings.
For our binarized embeddings, we thus firstly compute {$\mathcal{L}^{std}_{BPR}$} that calculates BPR loss (similar to Equation~(\ref{eq:hd-bpr})) with the student predictions from Equation~(\ref{eq:score}) as:
\begin{equation}
\mathcal{L}^{std}_{BPR} = -\sum_{u \in \mathcal{U}} \sum_{i\in \mathcal{N}(u)} \ln \sigma(\widehat{y}^{\,std}_{u,i} - \widehat{y}^{\,std, -}_{u,i}),
\end{equation}
where $\widehat{y}^{\,std}_{u,i}$ and $\widehat{y}^{\,std, -}_{u,i}$ are based on our binarized embedding version. 
Then we combine $\mathcal{L}^{std}_{BPR}$ with our dual inference distillation losses $\mathcal{L}_{ID_1}$ and $\mathcal{L}_{ID_2}$. 
The objective function is formulated as follows:
\begin{equation}
\new
\mathcal{L} = \mathcal{L}^{std}_{BPR} + \mathcal{L}_{ID_1} + \mathcal{L}_{ID_2} + \lambda ||\Theta||_2^2, 
\end{equation}
where {$||\Theta||_2^2$} is the $L$2-regularizer of node embeddings parameterized by hyper-parameter {$\lambda$} to avoid over-fitting.
}

\subsection{Gradient Estimation}
\label{sec:gradient}
While the \textit{Straight-Through Estimator (STE)}~\cite{bengio2013estimating} allows for gradient flow during backpropagation, it can lead to inconsistent optimization directions between forward and backward propagation. This is because the integral of the constant 1 in STE results in a linear function, rather than the true $\sign(\cdot)$ function. To provide more accurate gradient estimation, we propose using an approximation of the \textit{Dirac delta function}~\cite{bracewell1986fourier} for gradient estimation in this work.

\begin{figure}[tp]
\begin{minipage}{1\textwidth}
\includegraphics[width=\linewidth]{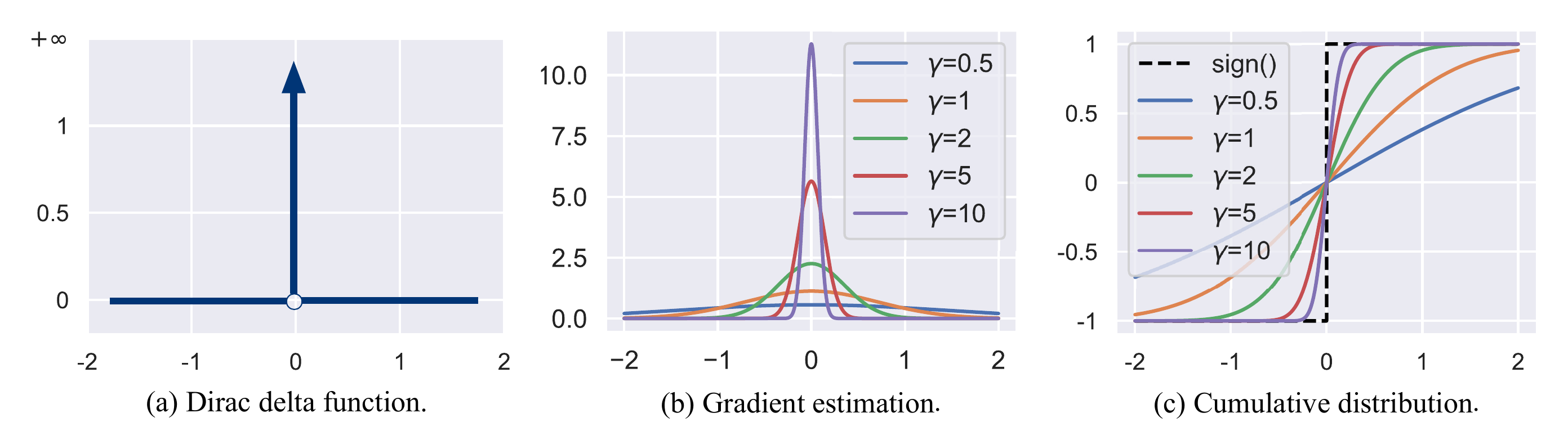}
\end{minipage} 
\caption{Illustration of our gradient estimation approach.}
\label{fig:gradient}
\end{figure}

Specifically, let {$u(\phi)$} represent the \textit{unit-step function}, also known as the Heaviside step function~\cite{step_function}, where {$u(\phi)$ $=$ $1$} for {$\phi > 0$} and {$u(\phi)$ $=$ $0$} otherwise. Clearly, we can translate the step function to the $\sign(\cdot)$ function using {$\sign(\phi) = 2u(\phi) - 1$}, which implies that {$\frac{\partial \sign(\phi)}{\partial \phi} = 2\frac{\partial u(\phi)}{\partial \phi}$}. For {$\frac{\partial u(\phi)}{\partial \phi}$}, it has been shown~\cite{bracewell1986fourier} that {$\frac{\partial u(\phi)}{\partial \phi} = 0$} when {$\phi \neq 0$} and {$\frac{\partial u(\phi)}{\partial \phi} = \infty$} when {$\phi = 0$}, which corresponds to the \textit{Dirac delta function}, also known as the unit impulse function {$\delta(\cdot)$}~\cite{bracewell1986fourier}, as depicted in Figure~\ref{fig:gradient}(a). However, directly using {$\delta(\cdot)$} for gradient estimation is impractical. A more feasible approach is to approximate {$\delta(\cdot)$} by introducing a zero-centered Gaussian probability density, as shown below:
\begin{equation}
 \delta(\phi) = \lim_{\beta \rightarrow \infty} \frac{|\beta|}{\sqrt{\pi}} \exp(-(\beta\phi)^2),
\end{equation}
which implies that: 
\begin{equation}
\label{eq:norm_gradient}
\frac{\partial \sign(\phi)}{\partial \phi} \doteq \frac{2\gamma}{\sqrt{\pi}} \exp(-(\gamma\phi)^2).
\end{equation}

As shown in Figure~\ref{fig:gradient}(b)-(c), the hyper-parameter {$\gamma$} controls the sharpness of the derivative curve for the $\sign(\cdot)$ approximation. Intuitively, our proposed gradient estimator aligns with the true gradient direction of $\sign(\cdot)$ during model optimization. This facilitates smooth quantization from continuous embeddings to quantized values, allowing for the estimation of evolving gradients across diverse input value ranges. As demonstrated in~\cref{exp:gradient}, our gradient estimator outperforms recent alternatives~\cite{gong2019differentiable,qin2020forward,darabi2018bnn,sigmoid,RBCN}. With all technical details covered, the pseudo-code of our model is provided in Algorithm~\ref{alg:bigear}.

\begin{algorithm}[t]
\caption{\new \model~algorithm.}
\label{alg:bigear}
\LinesNumbered  
\KwIn{Interaction graph; trainable embeddings {\footnotesize $\boldsymbol{v}_{\{\cdots\}}$}; hyper-parameters: {\footnotesize $L$, $\eta$, $\lambda$, $\lambda_1$, $\lambda_2$, $\gamma$. } }
\KwOut{Prediction function $\mathcal{F}(u,i)$.} 
$\mathcal{A}_u \gets \emptyset$, $\mathcal{A}_i \gets \emptyset$, $\mathcal{Q}_u \gets \emptyset$, $\mathcal{Q}_i \gets \emptyset$;\\
\While{\rm{\model~ not converge}}{
    \For{$l = 1, \cdots, L$}{
         $\boldsymbol{v}^{(l)}_u \gets\sum_{i\in \mathcal{N}(u)} \frac{1}{\sqrt{|\mathcal{N}(u)|\cdot|\mathcal{N}(i)|}}\boldsymbol{v}^{(l-1)}_i$, \\ 
         $\boldsymbol{v}^{(l)}_i\gets \sum_{u\in \mathcal{N}(i)} \frac{1}{\sqrt{|\mathcal{N}(i)|\cdot|\mathcal{N}(u)|}}\boldsymbol{v}^{(l-1)}_u$. \\
          \If{{with inference distillation}}{
              $\boldsymbol{q}_u^{(l)}\gets \sign\big(\boldsymbol{v}^{(l)}_u\big), \ \ \boldsymbol{q}_i^{(l)}\gets \sign\big(\boldsymbol{v}^{(l)}_i\big)$, \\
              $\alpha_u^{(l)} \gets \frac{||\boldsymbol{V}_u^{(l)}||_1}{d}$, $\alpha_i^{(l)} \gets \frac{||\boldsymbol{V}_i^{(l)}||_1}{d}$; \\
              Update ($\mathcal{A}_u$, $\mathcal{Q}_u$), ($\mathcal{A}_i$, $\mathcal{Q}_i$) with $\alpha_u^{(l)}\boldsymbol{q}_u^{(l)}$, $\alpha_i^{(l)}\boldsymbol{q}_i^{(l)}$; \\              
          }
        }
      {\new 
      $\widehat{y}^{\,tch}_{u,i} \gets \Big<\big|\big|_{l=0}^L w_l\boldsymbol{v}_u^{(l)}, \big|\big|_{l=0}^L w_l\boldsymbol{v}_i^{(l)}\Big>$; \\
      $\boldsymbol{v}_{i}^-$ $\gets$ pseudo-positive sample synthesis from the \text{synthesizer};  \\ 
      $\widehat{y}^{\,tch, -}_{u,i}$ $\gets$ $\big<\big|\big|_{l=0}^L w_l\boldsymbol{v}_u^{(l)}, \boldsymbol{v}_i^{-}\big>$; \\
      }
     
      \If{{ with inference distillation}}{
            $\boldsymbol{q}_u^{(0)} \gets \sign\big(\boldsymbol{v}^{(0)}_u\big), \ \ \boldsymbol{q}_i^{(0)} \gets \sign\big(\boldsymbol{v}^{(0)}_i\big)$, \\
              $\alpha_u^{(0)} \gets \frac{||\boldsymbol{V}_u^{(0)}||_1}{d}$, $\alpha_i^{(0)} \gets \frac{||\boldsymbol{V}_i^{(0)}||_1}{d}$; \\
              Update ($\mathcal{A}_u$, $\mathcal{Q}_u$), ($\mathcal{A}_i$, $\mathcal{Q}_i$) with $\alpha_u^{(0)}\boldsymbol{q}_u^{(0)}$, $\alpha_i^{(0)}\boldsymbol{q}_i^{(0)}$; \\
            $\widehat{y}^{\,std}_{u,i} =  \big<f(\mathcal{A}_u, \mathcal{Q}_u), f(\mathcal{A}_i, \mathcal{Q}_i)\big>$; \\ 
            $\{\widehat{y}^{tch,\,(l)}_{u,i}\}_{l=0, 1, \cdots, L} \gets \text{get score segments from } \widehat{y}^{\,tch}_{u,i}$; \\ 
             $\{\widehat{y}^{std,\,(l)}_{u,i}\}_{l=0, 1, \cdots, L} \gets \text{get score segments from }  \widehat{y}^{\,std}_{u,i}$; \\  
            {\new 
            $\mathcal{L}_{ID_1}$, $\mathcal{L}_{ID_2}$ $\gets$ calculate inference distillation loss for interacted and pseudo-positive training samples; \\
            $\boldsymbol{q}_{i}^-$ $\gets$ pseudo-positive sample synthesis from the synthesizer;  \\ 
            $\widehat{y}^{\,std, -}_{u,i}$ $\gets$ $\big<f(\mathcal{A}_u, \mathcal{Q}_u), \boldsymbol{q}_i^{-}\big>$; \\
            $\mathcal{L}_{BPR}^{std}$ $\gets$ calculate $\mathcal{L}_{BPR}^{std}$ with the positive score $\widehat{y}^{\,std}_{u,i}$ and the pseudo-positive one $\widehat{y}^{\,std, -}_{u,i}$.\\
            $\mathcal{L} \gets$ calculate $\mathcal{L}^{std}_{BPR}$, $\mathcal{L}_{ID_1}$, and $\mathcal{L}_{ID_2}$. \\ 

            }

      }
      \Else{
        {\new  $\mathcal{L} \gets$ calculate $\mathcal{L}^{tch}_{BPR}$ with the positive score $\widehat{y}^{\,tch}_{u,i}$ and the pseudo-positive one $\widehat{y}^{\,tch, -}_{u,i}$.} \\ 	
      }

       Optimize \model~with regularization;\\ 

}
\KwRet $\mathcal{F}$.\\
\end{algorithm}

\section{Model Analysis}
\label{sec:analysis}


\subsection{Magnification of Feature Uniqueness}
\label{sec:necessity}
We use user $u$ as an example for illustration, and the following analysis can be generalized to other nodes without loss of generality. 
Drawing on concepts from sensitivity analysis in statistics~\cite{koh2017understanding} and influence diffusion in social networks~\cite{xu2018representation}, we evaluate how the latent feature of a distant node $x$ impacts $u$'s representation segments before binarization (e.g., {$\boldsymbol{v}^{(l)}_u$}), assuming $x$ is a multi-hop neighbor of $u$.
We denote the \textbf{feature enrichment ratio} {$\mathbb{E}_{x \rightarrow u}^{(l)}$} as the L1-norm of Jacobian matrix {$\left[{\partial \boldsymbol{v}^{(l)}_u}/\ {\partial \boldsymbol{v}^{(0)}_{x_u}} \right]$}, by detecting the absolute influence of all fluctuation in entries of {$\boldsymbol{v}^{(0)}_{x}$} to {$\boldsymbol{v}^{(l)}_u$}, i.e., {$\mathbb{E}_{x\rightarrow u}^{(l)}$ = $\left|\left| \left[{\partial \boldsymbol{v}^{(l)}_u} /\ {\partial \boldsymbol{v}^{(0)}_{x}} \right] \right|\right|_1$}.
Focusing on a $l$-length path $h$ connected by the node sequence: {$x_h^l$, $x_h^{l-1}$, $\cdots$, $x_h^1$, $x_h^0$}, where {$x_h^l$ = $u$} and {$x_h^0$ = $x$}, we follow the chain rule to develop the derivative decomposition as:
\begin{equation}
\frac{\partial \boldsymbol{v}^{(l)}_u}{\partial \boldsymbol{v}^{(0)}_{x}}  = \sum_{h=1}^H
\left[\frac{\partial \boldsymbol{v}^{(l)}_{x_h^l}}{\partial \boldsymbol{v}^{(0)}_{x_h^0}} \right]_h
= \sum_{h=1}^H \prod_{k=l}^1
\frac{1}{\sqrt{|\mathcal{N}(x^k_h)}|} \cdot \frac{1}{\sqrt{|\mathcal{N}(x^{k-1}_h)}|} \cdot \boldsymbol{I}
=  \sqrt{\frac{|\mathcal{N}(u)|}{|\mathcal{N}(x)|}} \sum_{h=1}^H \prod_{k=1}^l  \frac{1}{|\mathcal{N}(x^k_h)|}  \cdot \boldsymbol{I},
\end{equation}
where $H$ is the number of paths between $u$ and $x$ in total.
Since all factors in the computation chain are positive, then:
\begin{equation}
\mathbb{E}_{x\rightarrow u}^{(l)} = \left|\left| \left[\frac{\partial \boldsymbol{v}^{(l)}_u}{\partial \boldsymbol{v}^{(0)}_{x}} \right]\right|\right|_1 = d \cdot \sqrt{\frac{|\mathcal{N}(u)|}{|\mathcal{N}(x)|}} \cdot \sum_{h=1}^H \prod_{k=1}^l  \frac{1}{|\mathcal{N}(x^k_h)|}.
\end{equation}
Note that here the term {$\sum_{h=1}^H \prod_{k=1}^l 1/|\mathcal{N}(x^k_h)|$} is exactly the probability of the $l$-length random walk starting at $u$ that finally arrives at $x$, which can be interpreted as:
\begin{equation}
\label{eq:prob}
\mathbb{E}_{x\rightarrow u}^{(l)} \propto \frac{1}{\sqrt{|\mathcal{N}(x)|}} \cdot Prob(\text{$l$-step random walk from $u$ arrives at $x$}).
\end{equation}

\textbf{Magnification Effect of Feature Uniqueness.}
Equation~(\ref{eq:prob}) suggests that with equal probability of visiting adjacent neighbors, distant nodes with fewer connections (i.e., {$|\mathcal{N}(x)|$}) will exert a greater influence on user $u$'s features. More importantly, in practice, these $l$-hop neighbors often represent \textit{esoteric} and \textit{unique} entities with lower popularity. By aggregating intermediate information from varying depths of graph convolution, we can achieve a \textbf{feature magnification effect}, amplifying the influence of unique nodes within $L$ hops of graph exploration. This ultimately enhances $u$'s semantic richness across all embedding segments for quantization.

\subsection{Complexity Analysis}
\label{sec:complexity}
To discuss the feasibility of realistic deployment, we compare \model~with the best full-precision model, LightGCN~\cite{lightgcn}, as both are \textit{end-to-end} systems involving offline model training and online prediction.

\begin{table}[t]
\centering
\small
\caption{Training time complexity.}
\label{tab:time}
{
\begin{tabular}{c | c | c | c | c | c}
\toprule
  ~   & {LightGCN}   &{BiGeaR$_{tch}$} 	&{\new \model$_{tch}$} 	&{BiGeaR$_{std}$} &{\new \model$_{std}$}  \\
\midrule
\midrule
{Graph Normalization}           &    \multicolumn{3}{c|}{$O\big(2E\big)$}  &  \multicolumn{2}{c}{-}\\
\midrule[0.1pt]
{Conv. and Bin.}   &   \multicolumn{3}{c|}{$O\big(\frac{2SdE^2L}{B}\big)$}  & \multicolumn{2}{c}{$O\big(2Sd(\frac{E^2L}{B}+(L+1)E)\big)$} \\
\midrule[0.1pt]
{\new $\mathcal{L}_{ID_1}$ Loss}     &   \multicolumn{2}{c|}{-}  & {\new $O\big(MNd(L+1)(N+\overline{\mathcal{N}}\ln \overline{\mathcal{N}})\big)$}  &{\new -}  & {\new $O\big(S\overline{\mathcal{N}} d(L+1)E \big)$} \\
\midrule[0.1pt]
{\new $\mathcal{L}_{ID_2}$ Loss}\tablefootnote{BiGeaR implemented an early version of inference distillation that shares the same complexity with $\mathcal{L}_{ID_2}$ loss in \model.}    &   {-}   & \multicolumn{2}{c|}{\new $O\big(MNd(L+1)(N+R\ln R)\big)$}  & \multicolumn{2}{c}{\new $O\big(SRd(L+1)E \big)$} \\
\midrule[0.1pt]
{\new Sample Synthesis}    &   \multicolumn{2}{c|}{-}   & {\new $O\big(2SJd(L+1)E\big)$}   &{\new -} & {\new $O\big(2SJd(L+1)E\big)$} \\
\midrule[0.1pt]
{$\mathcal{L}_{BPR}$ Loss}    &    {$O\big(2SdE\big)$}      & \multicolumn{4}{c}{$O\big(2Sd(L+1)E\big)$}  \\
\midrule[0.1pt]
{Gradient Estimation}  & \multicolumn{3}{c|}{-}  & \multicolumn{2}{c}{$O\big(2Sd(L+1)E\big)$}\\
\bottomrule
\end{tabular}}
\end{table}

\textbf{Training Time Complexity.}
Let {$M$}, {$N$}, and {$E$} represent the number of users, items, and edges, respectively, and let {$S$} and {$B$} denote the number of epochs and batch size. We use \model$_{tch}$ and \model$_{std}$ to refer to the pre-training version and the binarized version of our model.
As we can observe from Table~\ref{tab:time}, 
(1) both \model$_{tch}$ and \model$_{std}$ have asymptotically similar graph convolution complexity as LightGCN, with \model$_{std}$ incurring an additional {$O(2Sd(L+1)E)$} complexity due to binarization. 
{\new
(2) For {$\mathcal{L}_{ID_1}$}, based on the trained full-precision embeddings from \model$_{tch}$, we first compute the layer-wise prediction scores with time complexity of {$O(MNd(L+1))$}, followed by ranking the interacted items with {$O(N + \overline{\mathcal{N}} \ln \overline{\mathcal{N}})$}, where $\overline{\mathcal{N}}$ denotes the average number of these interacted items. The layer-wise inference distillation to \model$_{std}$ requires {$O(S\overline{\mathcal{N}} d(L+1)E)$}.
(3) Similarly, for {$\mathcal{L}_{ID_2}$}, the time complexity is {$O(MNd(L+1)(N+R\ln R))$} for full-precision embeddings and {$O(SRd(L+1)E)$} for binarized embeddings.
(4) For pseudo-positive sample synthesis, for each positive item in training, we use $J$ additional {\rev non-interacted} items as candidates to initialize the synthesis process, denoted as $\mathcal{E}_{can}$. This results in a time complexity $J$ times that of $\mathcal{L}_{BPR}$, with $J \leq 20$ as reported in Table~\ref{tab:hyperparameter}. 
Thus, compared to the convolution operation, the complexity of pseudo-positive sample synthesis is acceptable. To avoid the \textit{over-smoothing} issue~\cite{li2019deepgcns,li2018deeper}, we generally limit {$L \leq 4$}.
}
(5) To estimate the gradients for \model$_{std}$, it takes {$O(2Sd(L+1)E)$} for all training samples.

\textbf{Memory Overhead and Prediction Acceleration.}
We evaluate the memory footprint of embedding tables used for online prediction. 
As we can observe from the results in Table~\ref{tab:prediction}:
(1) Theoretically, the ratio of our model's embedding size to LightGCN's is {$\frac{32d}{(L+1)(32+d)}$}. Typically, with {$L \leq 4$} and {$d \geq 64$}, our model achieves at least a 4$\times$ reduction in space usage.
(2) In terms of prediction time, we compare the number of binary operations (\#BOP) and floating-point operations (\#FLOP) between our model and LightGCN. We find that \model~significantly reduces floating-point computations (e.g., multiplications) by replacing them with more efficient bitwise operations.

\begin{table}[t]
\centering

\caption{Complexity of space cost and online prediction.}
\label{tab:prediction}
{
\begin{tabular}{c | c | c | c}
\toprule
  ~          & {Embedding size}   &  {\#FLOP}	& {\#BOP}     	\\
\midrule
\midrule
 {LightGCN}      & {$O\big(32(M+N)d\big)$}       &   {$O\big(2MNd\big)$}      &   {-}       	\\
\midrule[0.1pt]
{\model}       & {$O\big((M+N)(L+1)(32+d)\big)$}     & {$O\big(4MN(L+1)\big)$}            & {$O\big(2MN(L+1)d\big)$}   	\\
\bottomrule
\end{tabular}}
\end{table}

\definecolor{high1}{RGB}{172, 226, 225}
\definecolor{high}{RGB}{218, 255, 251}
\newcommand{\fst}{\cellcolor{high1}}
\newcommand{\snd}{\cellcolor{high}}
\def\drop{\color{darkgray}}

\section{Experimental Results}
\label{sec:exp}
We evaluate our model on Top-K recommendation task with the aim of answering the following research questions:
\begin{itemize}[leftmargin=*]
{\new
\item \textbf{RQ1.} How does \model~ perform compared to state-of-the-art full-precision and quantization-based models?
}
\item \textbf{RQ2.} What is the practical resource consumption of \model?

{\new
\item \textbf{RQ3.} How do proposed components affect \model~performance?
}
\item \textbf{RQ4.} How do other hyper-parameter settings affect \model~performance?
\end{itemize} 

\subsection{Experimental Setups}
\subsubsection{\textbf{Datasets.}}
To guarantee the fair comparison, we directly use five experimented datasets (including the training/test splits) from: MovieLens\footnote{\url{https://grouplens.org/datasets/movielens/1m/}}~\cite{hashgnn,he2016fast,chen2021modeling,chen2021attentive}, Gowalla\footnote{\url{https://github.com/gusye1234/LightGCN-PyTorch/tree/master/data/gowalla}}~\cite{ngcf,hashgnn,zhang2024influential,dgcf}, Pinterest\footnote{\url{https://sites.google.com/site/xueatalphabeta/dataset-1/pinterest_iccv}}~\cite{geng2015learning,hashgnn}, Yelp2018\footnote{\url{https://github.com/gusye1234/LightGCN-PyTorch/tree/master/data/yelp2018}}~\cite{ngcf,dgcf,lightgcn}, and Amazon-Book\footnote{\url{https://github.com/gusye1234/LightGCN-PyTorch/tree/master/data/amazon-book}}~\cite{ngcf,dgcf,chen2023topological,chen2024shopping}.
Dataset statistics are reported in Table~\ref{tab:datasets}.
Specifically,
\begin{itemize}[leftmargin=*]

\item \textbf{MovieLens}~\cite{hashgnn,he2016fast,chen2021modeling,chen2021attentive} is a widely used benchmark for movie recommendation. Following the setup in~\cite{hashgnn,he2016fast,chen2021modeling}, we define $y_{u,i} = 1$ if user $u$ has given an explicit rating to item $i$, and $y_{u,i} = 0$ otherwise. In this work, we use the MovieLens-1M dataset split.

\item \textbf{Gowalla}~\cite{ngcf,hashgnn,lightgcn,dgcf} is a check-in dataset~\cite{liang2016modeling} from Gowalla, where users share their locations through check-ins. To ensure data quality, we follow~\cite{ngcf,hashgnn,lightgcn,dgcf} and filter users and items with at least 10 interactions.

\item \textbf{Pinterest}~\cite{geng2015learning,hashgnn} is an implicit feedback dataset used for image recommendation~\cite{geng2015learning}. Users and images are represented as a graph, with edges denoting user pins on images. Each user has a minimum of 20 edges.

\item \textbf{Yelp2018}~\cite{ngcf,dgcf,lightgcn} is sourced from the 2018 Yelp Challenge, where local businesses such as restaurants are considered items. Similar to~\cite{ngcf,dgcf,lightgcn}, we retain users and items with over 10 interactions.

\item \textbf{Amazon-Book}~\cite{ngcf,dgcf,lightgcn} is derived from the Amazon review collection for book recommendations~\cite{he2016ups}. Following the approach in~\cite{ngcf,lightgcn,dgcf}, we apply the 10-core setting to define the graph nodes.
\end{itemize}

\begin{table}[t]
\centering
\caption{The statistics of datasets.}
\label{tab:datasets}
{
\begin{tabular}{c | c  c  c  c  c}
\toprule 
             & {MovieLens}  & {Gowalla}   & {Pinterest}  &  {Yelp2018} & {Amazon-Book}\\
\midrule
\midrule[0.1pt]
    {\#Users}  & {6,040}   & {29,858}   & {55,186}   & {31,668}  &{52,643}  \\ 
    {\#Items}  & {3,952}   & {40,981}   & {9,916}    & {38,048}  &{91,599}  \\
\midrule[0.1pt]
    {\#Interactions} & {1,000,209} & {1,027,370} & {1,463,556} & {1,561,406} & {2,984,108} \\
    {Density} & {0.041902} & {0.000840} & {0.002675} & {0.001296} & {0.000619} \\
\bottomrule
\end{tabular}}
\end{table}

\subsubsection{\textbf{Competing Methods.}}
We evaluate the following recommender models: (1) 1-bit quantization-based methods, including graph-based models (GumbelRec~\cite{gumbel1,gumbel2}, HashGNN~\cite{hashgnn}) and general model-based approaches (LSH~\cite{lsh}, HashNet~\cite{hashnet}, CIGAR~\cite{kang2019candidate}); and (2) full-precision models, including neural-network-based (NeurCF~\cite{neurcf}) and graph-based models (NGCF~\cite{ngcf}, DGCF~\cite{dgcf}, LightGCN~\cite{lightgcn}). Specifically,
\begin{itemize}[leftmargin=*]
\item \textbf{LSH}~\cite{lsh} is a classic hashing method used for approximating similarity search in high-dimensional data. We adapt it for Top-K recommendation following the approach in~\cite{hashgnn}.

\item \textbf{HashNet}~\cite{hashnet} is a state-of-the-art deep hashing method, originally designed for multimedia retrieval. We use the same adaptation strategy from~\cite{hashgnn} for recommendation tasks.

\item \textbf{CIGAR}~\cite{kang2019candidate} is a hashing-based method for fast item candidate generation, followed by complex full-precision re-ranking. We use its quantization component for a fair comparison.

\item \textbf{GumbelRec} is a variant of our model that implements Gumbel-softmax for categorical variable quantization~\cite{gumbel1,gumbel2}, utilizing the Gumbel-softmax trick to replace the $\sign(\cdot)$ function for embedding binarization.

\item \textbf{HashGNN}~\cite{hashgnn} is a state-of-the-art, end-to-end 1-bit quantization recommender system. \textbf{HashGNN$_h$} refers to its standard \textit{hard encoding} version, while \textbf{HashGNN$_s$} represents a relaxed version that replaces some quantized digits with full-precision values.

\item \textbf{NeurCF}~\cite{neurcf} is a classical neural-network-based recommender system designed to capture non-linear user-item interactions for collaborative filtering.

\item \textbf{NGCF}~\cite{ngcf} is a state-of-the-art graph-based collaborative filtering model, which closely follows the structure of standard GCN~\cite{kipf2016semi}.

\item \textbf{DGCF}~\cite{dgcf} is a recent graph-based recommender model that disentangles user intents for improved Top-K recommendations.

\item \textbf{LightGCN}~\cite{lightgcn} is a highly efficient GCN-based recommender system with a simplified model architecture that delivers state-of-the-art performance.

\newxn{
\item \textbf{BiGeaR}~\cite{chen2022learning} is the most recent graph-based representation binarization model for recommendation, serving as a predecessor to this work. 
}
\end{itemize}

We exclude earlier quantization-based recommendation models, such as CH~\cite{liu2014collaborative}, DiscreteCF~\cite{zhang2016discrete}, and DPR~\cite{zhang2017discrete}, as well as full-precision solutions like GC-MC~\cite{berg2017graph} and PinSage~\cite{pinsage}, primarily because the competing models we include~\cite{kang2019candidate,lightgcn,ngcf,neurcf} have already demonstrated superiority over them.

\subsubsection{\textbf{Evaluation Metric.}}
For evaluating Top-K recommendation, we use two widely adopted metrics: Recall@K and NDCG@K, to assess the model's recommendation capability.

\subsubsection{\textbf{Implementation Details.}}
\label{sec:hyper_setting}
Our model is implemented in Python 3.7 using PyTorch 1.14.0, with non-distributed training. The experiments are conducted on a Linux machine equipped with 1 NVIDIA V100 GPU, 4 Intel Core i7-8700 CPUs, and 32 GB of RAM at 3.20GHz. For all baseline models, we use the officially reported hyper-parameter settings or perform a grid search for hyper-parameters if without recommended configurations. 
The embedding dimension is searched over \{$32, 64, 128, 256, 512, 1024$\}. 
The learning rate $\eta$ is tuned within \{$10^{-4}, 10^{-3}, 10^{-2}$\}, and the $L2$ regularization coefficient $\lambda$ is tuned among \{$10^{-6}, 10^{-5}, 10^{-4}, 10^{-3}$\}. All models are initialized and optimized using the default normal initializer and the Adam optimizer~\cite{adam}. We report all hyper-parameters in Table~\ref{tab:hyperparameter} to ensure reproducibility.

\begin{table}[thb]
\centering
\caption{Hyper-parameter settings for the five datasets.}
\label{tab:hyperparameter}
{\newxn
\begin{tabular}{c | c  c  c  c  c}
\toprule
                  & MovieLens   & Gowalla    & Pinterest     & Yelp2018  & Amazon-Book\\
\midrule 
\midrule
  $B$               &  2048       &   2048     & 2048        &  2048   &  2048  \\
  $d$         &  256        &   256      &  256    & 256  & 256   \\
  $\eta$            & $1\times10^{-3}$    & $1\times10^{-3}$  & $5\times10^{-4}$      & $5\times10^{-4}$        & $5\times10^{-4}$      \\
  $\lambda$         & $1\times10^{-4}$    & $5\times10^{-5}$  & $1\times10^{-4}$      & $1\times10^{-4}$        & $1\times10^{-6}$      \\
  $\lambda_1$   & 1 & 1 & 1 & 1 & 1\\
  $\lambda_2$   &  0.1 & 0.1 & 0.1 & 0.1 & 0.1 \\
  \newxn $J$           & 8 & 4 & 20 & 20 & 2 \\
  \newxn $c$           & 1 & 1 & 1 & 1 & 0.01 \\
  \newxn $R$           & 50 & 25 & 25 & 25 & 25 \\
  $\gamma$      & 1  & 1 & 1 & 1 & 1\\
  $L$       & 2 & 2 & 2 & 2 & 2 \\
\bottomrule
\end{tabular}}
\end{table}

\begin{table*}[tbh]
\centering
\small
  \caption{\new Performance comparison (the wavelines and underlines represent the best-performing full-precision and quantization-based models). We use Gain$^*$ to denote \model's performance improvement over BiGeaR.}
  \label{tab:top20}
  \setlength{\tabcolsep}{0.2mm}{
  \begin{tabular}{c|c c|c c|c c|c c|c c} 
    \toprule
    \multirow{2}*{Model} & \multicolumn{2}{c|}{MovieLens (\%)} & \multicolumn{2}{c|}{Gowalla (\%)} & \multicolumn{2}{c|}{Pinterest (\%)} & \multicolumn{2}{c|}{Yelp2018 (\%)} &  \multicolumn{2}{c}{Amazon-Book (\%)} \\
        ~ & Recall@20 & NDCG@20  & Recall@20 & NDCG@20  & Recall@20  & NDCG@20  & Recall@20  & NDCG@20  & Recall@20  & NDCG@20 \\
    \midrule
    \midrule
    NeurCF           & {21.40} $\pm$ {\footnotesize 1.51}   & {37.91} $\pm$ {\footnotesize 1.14}   & {14.64} $\pm$ {\footnotesize 1.75}  & {23.17} $\pm$ {\footnotesize 1.52} & {12.28} $\pm$ {\footnotesize 1.88} & {13.41} $\pm$ {\footnotesize 1.13}   & {4.28} $\pm$ {\footnotesize 0.71}   & {7.24} $\pm$ {\footnotesize 0.53}  &{3.49} $\pm$ {\footnotesize 0.75}  &{6.71} $\pm$ {\footnotesize 0.72} \\
    NGCF             & {24.69} $\pm$ {\footnotesize 1.67}  & {39.56} $\pm$ {\footnotesize 1.26}   & {16.22} $\pm$ {\footnotesize 0.90}  & {24.18} $\pm$ {\footnotesize 1.23}  & {14.67} $\pm$ {\footnotesize 0.56}  & {13.92} $\pm$ {\footnotesize 0.44} & {5.89} $\pm$ {\footnotesize 0.35} & {9.38} $\pm$ {\footnotesize 0.52}   &{3.65} $\pm$ {\footnotesize 0.73} &{6.90} $\pm$ {\footnotesize 0.65} \\
    DGCF             & {25.28} $\pm$ {\footnotesize 0.39}  & {45.98} $\pm$ {\footnotesize 0.58}   & {18.64} $\pm$ {\footnotesize 0.30}  & {25.20} $\pm$ {\footnotesize 0.41} & {\uwave{15.52}} $\pm$ {\footnotesize 0.42} & {\uwave{16.51}} $\pm$ {\footnotesize 0.56} & {6.37} $\pm$ {\footnotesize 0.55}  & {11.08} $\pm$ {\footnotesize 0.48}   &{4.32} $\pm$ {\footnotesize 0.34} &{7.73} $\pm$ {\footnotesize 0.27}\\
    LightGCN          & {\uwave{26.28}}  $\pm$ {\footnotesize 0.20}  & {\uwave{46.04}} $\pm$ {\footnotesize 0.18}   & {\uwave{19.02}}  $\pm$ {\footnotesize 0.19}  & {\uwave{25.71}} $\pm$ {\footnotesize 0.25}  & {15.33} $\pm$ {\footnotesize 0.28}  & {16.29} $\pm$ {\footnotesize 0.24}  & {\uwave{6.79}} $\pm$ {\footnotesize 0.31}  & {\uwave{12.17}} $\pm$ {\footnotesize 0.27} &{\uwave{4.84}} $\pm$ {\footnotesize 0.09}  &{\uwave{8.11}} $\pm$ {\footnotesize 0.11} \\
    \midrule[0.1pt]
    \rowcolor{high} 
    BiGeaR          &{25.57} $\pm$ {\footnotesize 0.33} &{45.56} $\pm$ {\footnotesize 0.31} &{18.36} $\pm$ {\footnotesize 0.14} &{24.96} $\pm$ {\footnotesize 0.17} &{15.57} $\pm$ {\footnotesize 0.22} &{16.83} $\pm$ {\footnotesize 0.46} &{6.47} $\pm$ {\footnotesize 0.14} &{11.60} $\pm$ {\footnotesize 0.18} &{4.68} $\pm$ {\footnotesize 0.11} &{8.12} $\pm$ {\footnotesize 0.12} \\ 
    \rowcolor{high} 
    \textbf{Capability} &{97.30\%} &{98.96\%} &{96.53\%} &{97.08\%} &{100.32\%} &{101.94\%} &{95.29\%} &{95.32\%} &{96.69\%} &{100.12\%} \\
    \rowcolor{high1} 
    {\newxn \textbf{\model}}  &{\newxn \textbf{26.52}} $\pm$ {\newxn \footnotesize 0.03} &{\newxn \textbf{46.82}} $\pm$ {\newxn \footnotesize 0.08} &{\newxn \textbf{18.66}} $\pm$ {\newxn \footnotesize 0.02} &{\newxn \textbf{25.23}} $\pm$ {\newxn \footnotesize 0.04} &{\newxn \textbf{16.40}} $\pm$ {\newxn \footnotesize 0.01} &{\newxn \textbf{17.43}} $\pm$ {\newxn \footnotesize 0.02} &{\newxn \textbf{6.80}} $\pm$ {\newxn \footnotesize 0.01} &{\newxn \textbf{12.02}} $\pm$ {\newxn \footnotesize 0.02} &{\newxn \textbf{5.16}} $\pm$ {\newxn \footnotesize 0.01} &{\newxn \textbf{8.78}} $\pm$ {\newxn \footnotesize 0.03}  \\   
    \rowcolor{high1} 
    \newxn \textbf{Capability} &{\newxn 100.91\%} &{\newxn 101.69\%}  &{\newxn 98.11\%} &{\newxn 98.13\%} &{\newxn 105.67\%} &{\newxn 105.57\%} &{\newxn  100.15\%} &{\newxn 98.77\%} &{\newxn 106.61\%} &{\newxn 108.26\%} \\
    \midrule 
    LSH             & {11.38} $\pm$ {\footnotesize 1.23}  & {14.87} $\pm$ {\footnotesize 0.76}   & {8.14} $\pm$ {\footnotesize 0.98}   & {12.19} $\pm$ {\footnotesize 0.86} & {7.88} $\pm$ {\footnotesize 1.21}  & {9.84} $\pm$ {\footnotesize 0.90}   & {2.91} $\pm$ {\footnotesize 0.51}   & {5.06} $\pm$ {\footnotesize 0.67} &{2.41}	$\pm$ {\footnotesize 0.95} &{4.39}	$\pm$ {\footnotesize 1.16} \\
    HashNet         & {15.43} $\pm$ {\footnotesize 1.73} & {24.78} $\pm$ {\footnotesize 1.50}   & {11.38} $\pm$ {\footnotesize 1.25}   & {16.50} $\pm$ {\footnotesize 1.42} & {10.27}  $\pm$ {\footnotesize 1.48} & {11.64} $\pm$ {\footnotesize 0.91}  & {3.37}  $\pm$ {\footnotesize 0.78}  & {7.31} $\pm$ {\footnotesize 1.16}  &{2.86}	$\pm$ {\footnotesize 1.51}	&{4.75}	$\pm$ {\footnotesize 1.33} \\
    CIGAR & {14.84} $\pm$ {\footnotesize 1.44} & {24.63} $\pm$ {\footnotesize 1.77}   & {11.57} $\pm$ {\footnotesize 1.01}   & {16.77} $\pm$ {\footnotesize 1.29} & {10.34}  $\pm$ {\footnotesize 0.97} & {11.87} $\pm$ {\footnotesize 1.20}  & {3.65}  $\pm$ {\footnotesize 0.90}  & {7.87} $\pm$ {\footnotesize 1.03}  &{3.05}  $\pm$ {\footnotesize 1.32}  &{4.98} $\pm$ {\footnotesize 1.24} \\
    GumbelRec       & {16.62} $\pm$ {\footnotesize 2.17} & {29.36} $\pm$ {\footnotesize 2.53} & {12.26}  $\pm$ {\footnotesize 1.58} & {17.49} $\pm$ {\footnotesize 1.08} & {10.53} $\pm$ {\footnotesize 0.79} & {11.86} $\pm$ {\footnotesize 0.86} & {3.85} $\pm$ {\footnotesize 1.39} & {7.97} $\pm$ {\footnotesize 1.59} &{2.69} $\pm$ {\footnotesize 0.55} &{4.32}  $\pm$ {\footnotesize 0.47} \\
    HashGNN$_h$     & {14.21} $\pm$ {\footnotesize 1.67} & {24.39} $\pm$ {\footnotesize 1.87}  & {11.63} $\pm$ {\footnotesize 1.47}  & {16.82} $\pm$ {\footnotesize 1.35}  & {10.15} $\pm$ {\footnotesize 1.43} & {11.96}  $\pm$ {\footnotesize 1.58} & {3.77}  $\pm$ {\footnotesize 1.02} & {7.75} $\pm$ {\footnotesize 1.39} &{3.09}	$\pm$ {\footnotesize 1.29} &{5.19} $\pm$ {\footnotesize 1.03}	\\
    HashGNN$_s$     & {\underline{19.87}} $\pm$ {\footnotesize 0.93} & {\underline{37.32}} $\pm$ {\footnotesize 0.81}  & {\underline{13.45}} $\pm$ {\footnotesize 0.65}  & {\underline{19.12}} $\pm$ {\footnotesize 0.68} & {\underline{12.38}}  $\pm$ {\footnotesize 0.86} & {\underline{13.63}} $\pm$ {\footnotesize 0.75} & {\underline{4.86}} $\pm$ {\footnotesize 0.36} & {\underline{8.83}} $\pm$ {\footnotesize 0.27} &{\underline{3.34}} $\pm$ {\footnotesize 0.25} 	&{\underline{5.82}} $\pm$ {\footnotesize 0.24}	\\
    \midrule[0.1pt]
    \rowcolor{high} 
    \textbf{BiGeaR}  &{\textbf{25.57}} $\pm$ {\footnotesize 0.33} &{\textbf{45.56}} $\pm$ {\footnotesize 0.31} &{\textbf{18.36}} $\pm$ {\footnotesize 0.14} &{\textbf{24.96}} $\pm$ {\footnotesize 0.17} &{\textbf{15.57}} $\pm$ {\footnotesize 0.22} &{\textbf{16.83}} $\pm$ {\footnotesize 0.46} &{\textbf{6.47}} $\pm$ {\footnotesize 0.14} &{\textbf{11.60}} $\pm$ {\footnotesize 0.18} &{\textbf{4.68}} $\pm$ {\footnotesize 0.11} &{\textbf{8.12}} $\pm$ {\footnotesize 0.12} \\  
    \rowcolor{high}  
    \textbf{Gain}   &{28.69\%} &{22.08\%} &{36.51\%} &{30.54\%} &{25.77\%} &{23.48\%} &{33.13\%} &{31.37\%}  &{40.12\%} &{39.52\%} \\
    \rowcolor{high1} 
   {\newxn \textbf{\model}}  &{\newxn \textbf{26.52}} $\pm$ {\newxn \footnotesize 0.03} &{\newxn \textbf{46.82}} $\pm$ {\newxn \footnotesize 0.08} &{\newxn \textbf{18.66}} $\pm$ {\newxn \footnotesize 0.02} &{\newxn \textbf{25.23}} $\pm$ {\newxn \footnotesize 0.04} &{\newxn \textbf{16.40}} $\pm$ {\newxn \footnotesize 0.01} &{\newxn \textbf{17.43}} $\pm$ {\newxn \footnotesize 0.02} &{\newxn \textbf{6.80}} $\pm$ {\newxn \footnotesize 0.01} &{\newxn \textbf{12.02}} $\pm$ {\newxn \footnotesize 0.02} &{\newxn \textbf{5.16}} $\pm$ {\newxn \footnotesize 0.01} &{\newxn \textbf{8.78}} $\pm$ {\newxn \footnotesize 0.03}  \\   
    \rowcolor{high1} 
    \newxn \textbf{Gain} &{\newxn 33.47\%} &{\newxn 25.46\%}  &{\newxn 38.74\%} &{\newxn 31.96\%} &{\newxn 32.47\%} &{\newxn 27.88\%} &{\newxn 39.92\%} &{\newxn 36.13\%} &{\newxn 54.49\%} &{\newxn 50.86\%} \\
    \rowcolor{high1} 
    \new \textbf{Gain$^*$} &{\new 3.72\%} &{\new 2.77\%}  &{\new 1.63\%} &{\new 1.08\%} &{\new 5.33\%} &{\new 3.57\%} &{\new 5.10\%} &{\new 3.62\%} &{\new 10.26\%} &{\new 8.13\%}  \\
    \bottomrule
    \bottomrule
  \end{tabular}}
\end{table*}


\subsection{Performance Analysis (RQ1)}
We evaluate Top-K recommendation by varying K in \{20, 40, 60, 80, 100\}.
We summarize the Top@20 results in Table~\ref{tab:top20} for detailed comparison. We have the following observations:
{\new
\begin{itemize}[leftmargin=*]
\item \textbf{\model~achieves competitive performance compared to state-of-the-art full-precision recommender models.}  
(1) While BiGeaR generally outperforms most full-precision recommender models (excluding LightGCN) across five benchmarks, our \model~model further enhances performance, achieving competitive results.  
The reasons are twofold.  
First, as analyzed in~\cite{chen2022learning}, \model~captures various levels of interactive information across multiple depths of graph exploration, significantly enriching user-item representations for binarization.  
Second, we introduce two novel designs in \model with pseudo-positive sample learning enhancement: dual inference distillation and pseudo-positive sample synthesis, both of which are empirically effective for Top-K ranking and recommendation tasks.  
(2) Compared to the state-of-the-art LightGCN, \model~achieves 98–108\% of the performance capability \textit{w.r.t.} Recall@20 and NDCG@20 across all datasets.  
This demonstrates that \model's design effectively narrows the performance gap with full-precision models like LightGCN. Considering the space compression and inference acceleration advantages discussed later, we argue that this performance is satisfactory, particularly in resource-limited deployment scenarios.

\item \textbf{Compared to all binarization-based recommender models, \model~consistently delivers impressive and statistically significant performance improvements.}  
(1) Two conventional methods (LSH, HashNet) for general item retrieval tasks underperform compared to CIGAR, HashGNN, BiGeaR, and \model, indicating that direct model adaptations are too simplistic for Top-K recommendation.  
(2) Graph-based models, such as CIGAR, generally perform better because CIGAR combines neural networks with \textit{learning to hash} techniques for fast candidate generation, whereas graph-based models better explore multi-hop interaction subgraphs, directly simulating the high-order \textit{collaborative filtering} process for model learning.  
(3) Our improved \model~outperforms HashGNN by approximately 32\%–54\% and 25\%–51\% \textit{w.r.t.} Recall@20 and NDCG@20, respectively.
Considering the further improvements of 1.08\%$\sim$10.26\% over its predecessor BiGeaR, all these results demonstrate the effectiveness of our proposed binarization components.


\end{itemize}
}

\begin{table}[t]
\caption{Resource consumption on MovieLens dataset.}
\label{tab:consumption}
{
\begin{tabular}{c | c  c  c  c c}
\toprule
              & LightGCN   &  HashGNN$_h$ &  HashGNN$_s$  &  BiGeaR & {\new \model}\\
\midrule[0.1pt]
\midrule[0.1pt]
  $T_{train}\mathbin{/}${\#epcoch}     &   {4.91s}     &   {186.23s}  & {204.53s}    &   {(5.16+6.22)s} &{\new (11.23+14.50)s}  \\
\midrule[0.1pt]
   $T_{infer}\mathbin{/}${\#query}    & {32.54ms}    & {2.45ms}  & {31.76ms}   & {3.94ms} &{\new 3.96ms} \\
\midrule[0.1pt]
  $S_{ET}$    & {9.79MB}    & {0.34MB}  & {9.78MB}   & {1.08MB} &{\new 1.08MB}\\
\midrule
\midrule
  Recall@20    & {26.28\%}    & {14.21\%}  & {19.87\%}   & {25.57\%} &{\new 26.52\%}\\
\bottomrule
\end{tabular}}
\end{table}

\subsection{Resource Consumption Analysis (RQ2)}
We analyze the resource consumption in \textit{training}, \textit{online inference}, and \textit{memory footprint} by comparing to the best two competing models, i.e., LightGCN and HashGNN.
Due to the page limits, we report the empirical results of MovieLens dataset in Table~\ref{tab:consumption}.
\begin{enumerate}[leftmargin=*]
\item $T_{train}$: We set the batch size to $B=2048$ and the dimension size to $d=256$ for all models. We observe that HashGNN is significantly more time-consuming compared to LightGCN, BiGeaR, and \model. This is because HashGNN is built on the earlier GCN framework~\cite{graphsage}, whereas LightGCN, BiGeaR, and \model~use a more streamlined graph convolution architecture that eliminates operations such as self-connections, feature transformations, and nonlinear activations~\cite{lightgcn}.
{\newxn Furthermore, with BiGeaR takes around 11s per epoch for pre-training and quantization, \model~ needs 25s for these stages, both two models take slightly more yet asymptotically similar time cost with LightGCN, basically following the complexity analyses in~\cref{sec:complexity}.
}

\item $T_{infer}$: We randomly generate 1,000 queries for online prediction and perform experiments using vanilla NumPy\footnote{{\url{https://www.lfd.uci.edu/~gohlke/pythonlibs/}}} on CPUs. We observe that HashGNN$_s$ has a similar time cost to LightGCN. 
This is because, while HashGNN$_h$ purely binarizes the continuous embeddings, the relaxed version HashGNN$_s$ uses a Bernoulli random variable to determine the probability of replacing quantized digits with their original real values~\cite{hashgnn}. 
As a result, although HashGNN$_h$ can accelerate predictions using Hamming distance, HashGNN$_s$, which improves recommendation accuracy, relies on floating-point arithmetic. In contrast, both BiGeaR and \model~benefit from bitwise operations, running about 8$\times$ faster than LightGCN while maintaining similar performance on the MovieLens dataset.

\item $S_{ET}$: We only store the embedding tables required for online inference. As mentioned earlier, HashGNN$_s$ interprets embeddings using randomly selected real values, leading to increased space consumption. Unlike HashGNN$_s$, both BiGeaR and \model~can store binarized embeddings and their corresponding scalers separately, achieving a balanced trade-off between recommendation accuracy and storage efficiency.

\end{enumerate}

\subsection{Study of Layer-wise Quantization (RQ3.A)}
\label{sec:lw_exp}
\begin{figure}[tp]
\begin{minipage}{1\textwidth}
\includegraphics[width=\linewidth]{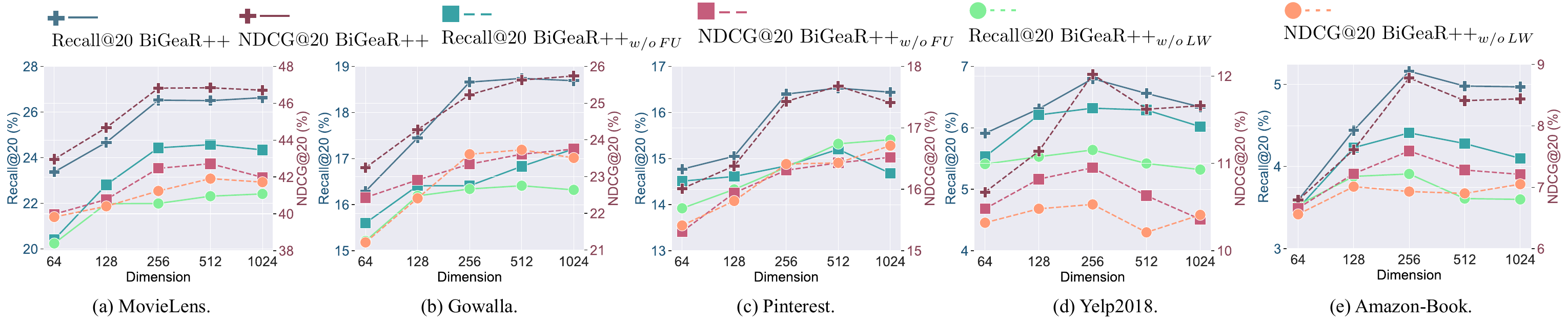}
\end{minipage} 
\caption{\new Study of graph layer-wise quantization on \model.}
\label{fig:dim_layer}
\end{figure}

To investigate the amplification of feature uniqueness in layer-wise quantization, we modify \model~and propose two variants, denoted as \model$_{w/o\,LW}$ and \model$_{w/o\,FU}$. 
We report the results in Figure~\ref{fig:dim_layer} by denoting Recall@20 and NDCG@20 in cold and warm colors, respectively. From these results, we have the following explanations.
\begin{itemize}[leftmargin=*]
\item First, \model$_{w/o\,LW}$ removes the layer-wise quantization and adopts the traditional approach of quantizing the final outputs from $L$ convolution iterations. We vary the dimension $d$ from 64 to 1024 while fixing the layer number $L=2$ for \model. 
{\new As the dimension size increases from 64 to 256, \model$_{w/o\,LW}$ shows a rapid improvement in both Recall@20 and NDCG@20 performance. However, when $d$ increases from 256 to 1024, the performance gains slow down or even decrease, likely due to overfitting. Therefore, with a moderate dimension size of $d=256$, \model~can achieve strong performance with reasonable computational complexity. 
}
\item Second, \model$_{w/o\,FU}$ omits the feature magnification effect by adopting the way used in HashGNN~\cite{graphsage,hashgnn} as:
\begin{equation}
\begin{matrix}
\boldsymbol{v}^{(l)}_x = \sum_{z\in \mathcal{N}(x)} \frac{1}{|\mathcal{N}(z)|}\boldsymbol{v}^{(l-1)}_z.
\end{matrix}
\end{equation}
As discussed in~\cref{sec:necessity}, this modification removes the ``magnification term'' in Equation~(\ref{eq:prob}), reducing it to a standard random walk for graph exploration. While \model$_{w/o\,FU}$ follows similar trends to \model~as the embedding dimension increases, its overall performance across all five datasets is inferior to \model. This confirms the importance of \model's feature magnification in enhancing unique latent feature representations, which enrich user-item embeddings and improve Top-K recommendation performance.
\end{itemize}

\subsection{\newxn Study of Dual Inference Distillation (RQ3.B)} 

\subsubsection{\textbf{Effect of Layer-wise Distillation.}}
We evaluate the effectiveness of our inference distillation by introducing several ablation variants, specifically \textsl{w/o ID} and \textsl{endID}. The \textsl{w/o ID} variant completely removes inference distillation during model training, while \textsl{endID} reduces the original layer-wise distillation to only distill information from the final layer of graph convolution. As shown in Table~\ref{tab:distillation}, both \textsl{w/o ID} and \textsl{endID} result in significant performance degradation. Additionally, the performance gap between \textsl{endID} and \model~demonstrates the effectiveness of applying inference distillation in a layer-wise manner to achieve further performance improvements.

\begin{table}[t]
\caption{\newxn Learning dual inference distillation.}
\label{tab:distillation}
{\newxn
\begin{tabular}{c |c c|c c|c c|c c|c c}
\toprule
 \multirow{2}*{Variant} & \multicolumn{2}{c|}{MovieLens} & \multicolumn{2}{c|}{Gowalla} & \multicolumn{2}{c|}{Pinterest} & \multicolumn{2}{c|}{Yelp2018} & \multicolumn{2}{c}{Amazon-book} \\
               ~  & R@20 & N@20 & R@20 & N@20 & R@20 & N@20 & R@20 & N@20 & R@20 & N@20\\
\midrule
\midrule
    \new \multirow{2}*{\textsl{w/o ID}}    &{\newxn 25.96}& {\newxn 46.17}   & {\newxn 18.17}& {\newxn 24.88}  & {\newxn 16.20}& {\newxn 17.29}  & {\newxn 6.62} & {\newxn 11.83} & {\newxn 4.66} & {\newxn 8.09}\\
~        &\textit{\drop {-2.11\%}}  &\textit{\drop {-1.39\%}}  &\textit{\drop {-2.63\%}}  &\textit{\drop {-1.39\%}}  &\textit{\drop {-1.22\%}}  &\textit{\drop {-0.80\%}}  &\textit{\drop {-2.65\%}}  &\textit{\drop {-1.58\%}} &\textit{\drop {-9.69\%}}  &\textit{\drop  {-7.86\%}}\\
  \midrule[0.1pt]
    \multirow{2}*{\textsl{endID}}    &{26.14}& {45.13}   & {18.44}& {25.10}  & {16.33}& {17.27}  & {6.76} & {11.98} & {4.82} & {8.29}\\
~           &\textit{\drop {-1.43\%}}   &\textit{\drop {-3.61\%}}   &\textit{\drop {-1.18\%}}   &\textit{\drop {-0.52\%}}   &\textit{\drop {-0.43\%}}  &\textit{\drop {-0.92\%}}  &\textit{\drop {-0.59\%}}   &\textit{\drop {-0.33\%}}   &\textit{\drop {-6.59\%}}   &\textit{\drop {-5.58\%}} \\
\midrule[0.1pt]
  \multirow{2}*{\textsl{KLD}}    &{26.08}& {46.19}   & {18.19}& {24.97}  & {15.44}& {17.06}  & {6.45} & {11.30} & {4.63} & {7.96}\\
  ~         &\textit{\drop {-1.66\%}}   &\textit{\drop {-1.34\%}}   &\textit{\drop {-2.52\%}}   &\textit{\drop {-1.03\%}}   &\textit{\drop {-5.85\%}}   &\textit{\drop {-2.12\%}}   &\textit{\drop {-5.15\%}}   &\textit{\drop {-5.99\%}}   &\textit{\drop {-10.27\%}}   &\textit{\drop {-9.34\%}} \\
  \midrule[0.1pt]
  \new \multirow{2}*{\textsl{w/o ID$_1$}}    &{26.42}& {46.78}   & {18.65}& {25.19}  & {16.39}& {17.44}  & {6.77} & {12.02} & {5.15} & {8.77}\\
  ~        &\textit{\drop {-0.38\%}}  &\textit{\drop {-0.09\%}}  &\textit{\drop {-0.05\%}}  &\textit{\drop {-0.15\%}}  &\textit{\drop {-0.06\%}}  &\textit{\drop {-}}  &\textit{\drop {-0.44\%}}  &\textit{\drop {-}} &\textit{\drop {-0.19\%}}  &\textit{\drop  {-0.11\%}}\\
  \midrule[0.1pt]
  \new \multirow{2}*{\textsl{w/o ID$_2$}}    &{26.21}& {46.62}   & {18.37}& {24.98}  & {16.26}& {17.33}  & {6.67} & {11.89} & {5.00} & {8.70}\\
  ~        &\textit{\drop {-1.17\%}}  &\textit{\drop {-0.43\%}}  &\textit{\drop {-1.55\%}}  &\textit{\drop {-0.99\%}}  &\textit{\drop {-0.85\%}}  &\textit{\drop {-0.57\%}}  &\textit{\drop {-1.91\%}}  &\textit{\drop {-1.08\%}} &\textit{\drop {-3.10\%}}  &\textit{\drop  {-0.91\%}}\\
  \midrule[0.1pt]
{\textbf{\model} }  &\textbf{26.52}& \textbf{46.82}   & \textbf{18.66}& \textbf{25.23}  & \textbf{16.40}& \textbf{17.43}  & \textbf{6.80}& \textbf{12.02} & \textbf{5.16}& \textbf{8.78}\\
\bottomrule
\end{tabular}}
\end{table}

\subsubsection{\textbf{Conventional Knowledge Distillation.}}
To compare with the conventional method, we modify \model~by applying KL divergence to the layer-wise teacher and student logits, i.e., {$\boldsymbol{\widehat{y}}_u^{,tch, (l)}$} vs. {$\boldsymbol{\widehat{y}}_u^{\,std, (l)}$}, and refer to this variant as \textsl{KLD}. As shown in Table~\ref{tab:distillation}, employing the conventional knowledge distillation with KL divergence results in suboptimal performance. This occurs because KL divergence encourages the alignment of logit distributions between the teacher and student models, but fails to effectively capture users' relative preferences for items. In contrast, our proposed layer-wise inference distillation proves to be more effective for distilling ranking information.

{\newxn 
\subsubsection{\textbf{Effect of Our Dual Distillation Design.}}
To explore the individual impacts of extracting positive and pseudo-positive training samples on distillation performance, we design two ablation variants: \textit{w/o ID$_1$} and \textit{w/o ID$_2$}. 
In \textit{w/o ID$_1$}, the {$\mathcal{L}_{ID_1}$} term is removed, while \textit{w/o ID$_2$} removes {$\mathcal{L}_{ID_2}$}. 
As shown in Table~\ref{tab:distillation}, compared to removing the teacher's positive ranking information in \textit{w/o ID$_1$}, omitting the pseudo-positive sample information in \textit{w/o ID$_2$} results in a more pronounced degradation. 
These findings suggest that the supervisory signal from {\rev non-interacted} items is more influential than that of interacted ones. 
However, combining both signals allows the student binarized embeddings to learn from a more holistic distillation process, achieving optimal performance.

\subsection{Study of Pseudo-positive Embedding Sample Synthesis (RQ3.C)}
\subsubsection{\textbf{Effectiveness of Sample Synthesis in Training.}}

To verify the effectiveness of pseudo-positive sample synthesis, we designed two ablation variants: \textit{w/o SS-FP} and \textit{w/o SS-B}. 
The \textit{w/o SS-FP} variant disables the pseudo-positive sample synthesis during full-precision pre-training, while the \textit{w/o SS-B} variant excludes it during the binarization training phase. 
As presented in Table~\ref{tab:synthesis}, both variants show varying levels of performance degradation.
The results highlight the crucial role of contributing significantly to generating pseudo-positive samples, thereby improving the model's ability to distinguish between relevant and irrelevant information during both training stages.

\begin{table}[t]
\centering
\caption{\newxn Learning pseudo-positive sample synthesis.}
\label{tab:synthesis}
{\newxn 
\begin{tabular}{c |c c|c c|c c|c c|c c}
\toprule
 \multirow{2}*{Variant} & \multicolumn{2}{c|}{MovieLens} & \multicolumn{2}{c|}{Gowalla} & \multicolumn{2}{c|}{Pinterest} & \multicolumn{2}{c|}{Yelp2018} & \multicolumn{2}{c}{Amazon-book} \\
               ~  & R@20 & N@20 & R@20 & N@20 & R@20 & N@20 & R@20 & N@20 & R@20 & N@20\\
\midrule
\midrule
  \multirow{2}*{\textsl{w/o SS-FP}}    &{26.43}& {46.64}   & {18.37}& {24.91}  & {15.99}& {17.15}  & {6.64} & {11.69} & {4.79} & {8.17}\\
  ~        &\textit{\drop {-0.34\%}}  &\textit{\drop {-0.38\%}}  &\textit{\drop {-1.55\%}}  &\textit{\drop {-1.27\%}}  &\textit{\drop {-2.50\%}}  &\textit{\drop {-1.61\%}}  &\textit{\drop {-2.36\%}}  &\textit{\drop {-2.75\%}} &\textit{\drop {-7.17\%}}  &\textit{\drop  {-6.95\%}}\\
  \midrule[0.1pt]   
  \multirow{2}*{\textsl{w/o SS-B}}    &{26.28}& {46.60}   & {18.56}& {25.10}  & {16.09}& {17.14}  & {6.80} & {11.87} & {5.15} & {8.67}\\
  ~        &\textit{\drop {-0.90\%}}  &\textit{\drop {-0.47\%}}  &\textit{\drop {-0.54\%}}  &\textit{\drop {-0.52\%}}  &\textit{\drop {-1.89\%}}  &\textit{\drop {-1.66\%}}  &\textit{-}  &\textit{\drop {-1.25\%}} &\textit{\drop {-0.19\%}}  &\textit{\drop  {-1.25\%}}\\
  \midrule[0.1pt]
{\textbf{\model} }  &\textbf{26.52}& \textbf{46.82}   & \textbf{18.66}& \textbf{25.23}  & \textbf{16.40}& \textbf{17.43}  & \textbf{6.80}& \textbf{12.02} & \textbf{5.16}& \textbf{8.78}\\
\bottomrule
\end{tabular}}
\end{table}

\subsubsection{\textbf{Implementation of $\boldsymbol{\beta}^{(l)}$.}}
We evaluate the design of $\boldsymbol{\beta}^{(l)} = [{\beta}_{0}^{(l)}, {\beta}_{1}^{(l)}, \ldots, {\beta}_{d-1}^{(l)}]$ employed in \textit{positive mix-up} for pseudo-positive sample synthesis by comparing our approach with different state-of-the-art ones.
\begin{enumerate}
\item MixGCF~\cite{MixGCF2021} sets constant ${\beta}^{(l)}$ as ${\beta}^{(l)} \sim \mathcal{U}(0, 1) \ $ for \ $l = 0, 1, \ldots, L.$ 

\item DINS~\cite{DINS2023}, for user $u$, sets the ${\beta}^{(l)}$ as follows. Let $j^*$ and $i$ denote the hardest item in $\mathcal{E}_{can}$ and the positive item, respectively. $w'$ is a weight and $d = 0, 1, \ldots, d-1$.
We use $e_u^d$ to denote the $d$-th embedding element.
\begin{equation}
{\beta}_{d}^{(l)} = \frac{\exp(e^{d}_{u}\cdot e^{d}_{i})}{\exp(e^{d}_{u}\cdot e^{d}_{i}) + w' \cdot \exp(e^{d}_{u}\cdot e^{d}_{j^*})}    
\end{equation}

\end{enumerate}
As shown in Table~\ref{tab:beta} and Table~\ref{tab:datasets}, Implementation (1) performs better on sparser datasets (Gowalla, Yelp2018, and Amazon-Book), while Implementation (2) excels on denser datasets (MovieLens and Pinterest). 
Meanwhile, our implementation demonstrates balanced performance across different levels of sparsity, consistently performing at least sub-optimally on all datasets and emerging as the top model for Gowalla and Amazon-Book. These findings highlight the effectiveness of our proposed approach in pseudo-positive sample synthesis.

\begin{table}[t]
\centering
\caption{\new Implementation of \textbf{$\beta$} (bold font represents the best model and underline represents the second-best model).}
\label{tab:beta}
{\newxn
\begin{tabular}{c | c c | c c | c c | c c | c c }
\toprule
\multirow{2}*{Implementation} & \multicolumn{2}{c|}{MovieLens} & \multicolumn{2}{c|}{Gowalla} & \multicolumn{2}{c|}{Pinterest} & \multicolumn{2}{c|}{Yelp2018} & \multicolumn{2}{c}{Amazon-Book}\\
               ~  & R@20 & N@20 & R@20 & N@20 & R@20 & N@20 & R@20 & N@20 &R@20 & N@20\\
\midrule[0.1pt]
\midrule[0.1pt]
  (1)           &{26.34}  &{46.67} &{\underline{18.65}} &{\underline{25.19}} &{16.34} &{17.41} &{\textbf{6.82}} &{\textbf{12.04}} &{\underline{4.99}} &{\underline{8.54}} \\
  (2)           &{\textbf{26.79}}  &{\textbf{47.24}} &{18.47} &{25.01} &{\textbf{16.62}} &{\textbf{17.66}} &{6.79} &{12.01} &{4.84} &{8.46} \\   
  \midrule[0.1pt]   
  \textbf{\model}     &{\underline{26.52}}  &{\underline{46.82}} &{\textbf{18.66}} &{\textbf{25.23}} &{\underline{16.40}} &{\underline{17.43}} &{\underline{6.80}} &{\underline{12.02}} &{\textbf{5.16}} &{\textbf{8.78}} \\
\bottomrule
\end{tabular}}
\end{table}
}

{\rev
\subsubsection{\textbf{Size $J$ of Negative Items in Sample Synthesis.}}

\begin{figure}[tp]
\begin{minipage}{1\textwidth}
\includegraphics[width=\linewidth]{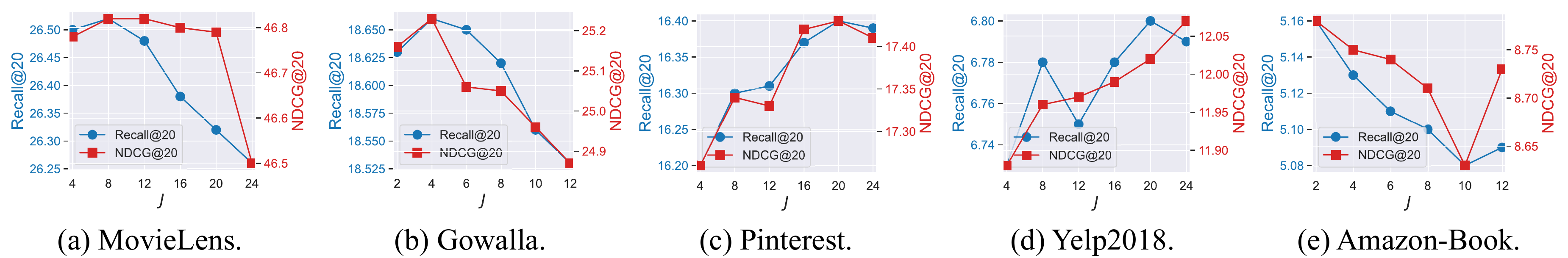}
\end{minipage} 
\caption{\rev Varying $J$ of negative items in sample synthesis.}
\label{fig:j}
\end{figure}

To validate the effect of $J$ settings in pseudo-positive sample synthesis, we conduct experiments across all datasets by varying $J$.
The results of Recall@20 and NDCG@20 are reported in Figure~\ref{fig:j}.
Generally, the setting of $J$ is empirically determined and varies across different datasets.
We take the MovieLens dataset as an example for explanation. 
We observe that lower $J$ value ($J$ $<$ 8) results in insufficient hard negative mining, while higher values ($J$ $>$ 8) introduce excessive noise from the uninformative negative samples, both scenarios leading to performance degradation. 
Therefore, for MovieLens dataset, a suitable configuration $J$ = 8 balances the information utilization and noise incorporation, and thus achieves the optimal performance.

}

\subsection{Study of Gradient Estimation (RQ3.D)}
\label{exp:gradient}
\begin{figure}[t]
\begin{minipage}{0.9\textwidth}
\includegraphics[width=\linewidth]{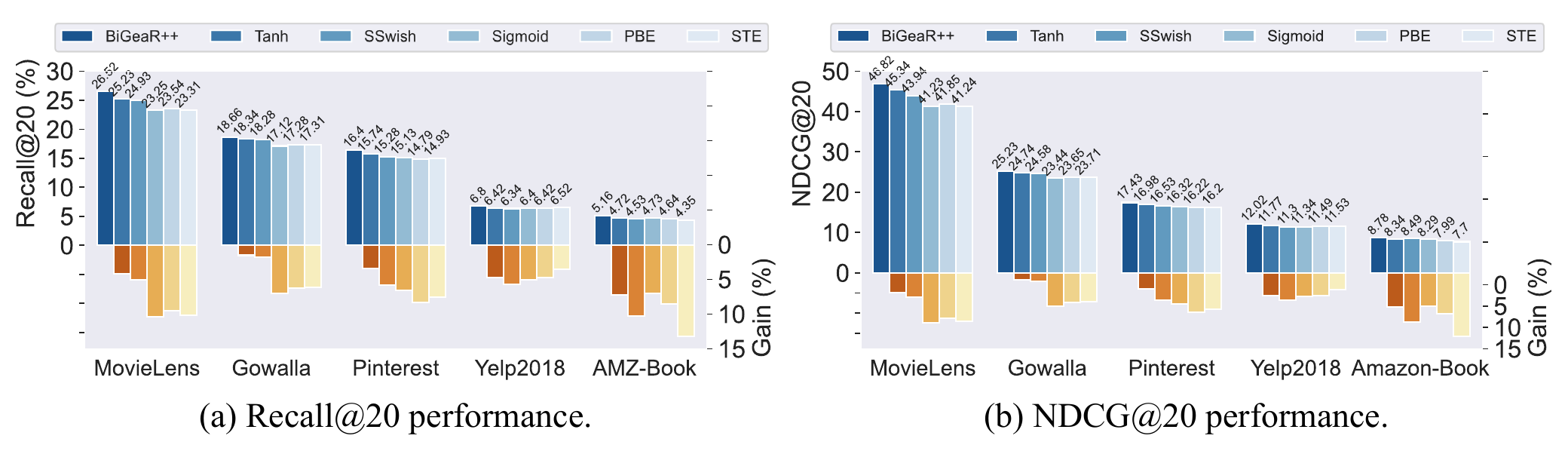}
\end{minipage} 
\caption{\new Gradient estimator comparison \textit{w.r.t.} Recall@20 and NDCG@20.}
\label{fig:quant_f}
\end{figure}

We evaluate our gradient estimation method by comparing it with several recently proposed estimators, including \textit{Tanh-like}~\cite{qin2020forward,gong2019differentiable}, \textit{SSwish}~\cite{darabi2018bnn}, \textit{Sigmoid}~\cite{sigmoid}, and the \textit{projected-based estimator} (PBE)~\cite{RBCN}, all implemented in \model. The Recall@20 results of these methods are presented in Figure~\ref{fig:quant_f}, along with the performance gains of our approach relative to these estimators.
We draw two key observations:
\begin{enumerate}[leftmargin=*]
\item Our method consistently outperforms the other gradient estimators. While these estimators rely on \textit{visually similar} functions, such as tanh($\cdot$), to approximate $\sign(\cdot)$, they lack a \textit{theoretical connection} to $\sign(\cdot)$, which can result in inaccurate gradient estimates. In contrast, as detailed in~\cref{sec:gradient}, our approach directly transforms the unit-step function $u(\cdot)$ into $\sign(\cdot)$ using the relation $\sign(\cdot)$ = 2$u(\cdot)$ - 1, enabling us to estimate the gradients of $\sign(\cdot)$ through the approximated derivatives of $u(\cdot)$. This method aligns both forward and backward propagation with the actual gradients of $\sign(\cdot)$, resulting in more accurate gradient estimation compared to previous methods.

\item The MovieLens dataset shows a larger performance gap between our method and the others compared to the last four datasets. This is attributed to the higher density of the MovieLens dataset, where the interaction-to-user-item ratio, $\frac{\#Interactions}{\#Users \cdot \#Items}$, is 0.0419—significantly higher than the corresponding ratios of {0.00084, 0.00267, 0.0013, 0.00062} for the other datasets. The higher interaction density reflects more complex user preferences and item interactions, placing greater demands on gradient estimation to learn accurate ranking information. As shown in Figure~\ref{fig:quant_f}, our method is particularly effective in handling these challenges in denser interaction graphs.
\end{enumerate}

\subsection{Study of Other Hyper-parameter Settings (RQ4)}

\subsubsection{\textbf{Implementation of Embedding Scaler ${\alpha}^{(l)}$}.}
\label{app:scaler}

We set the embedding scaler to be learnable (denoted as \textsl{LB}), with the results presented in Table~\ref{tab:scaler}. However, we observe that the learnable embedding scaler does not deliver the expected performance. This may be due to the absence of a direct mathematical constraint, which leads to an overly large parameter search space, making it difficult for stochastic optimization to find the optimal solution.

\begin{table}[t]
\centering
\caption{\new Implementation of embedding scaler ${\alpha}^{(l)}$.}
\label{tab:scaler}
{\new
\begin{tabular}{c |c c|c c|c c|c c|c c}
\toprule
 \multirow{2}*{Implementation} & \multicolumn{2}{c|}{MovieLens} & \multicolumn{2}{c|}{Gowalla} & \multicolumn{2}{c|}{Pinterest} & \multicolumn{2}{c|}{Yelp2018} & \multicolumn{2}{c}{Amazon-book} \\
               ~  & R@20 & N@20 & R@20 & N@20 & R@20 & N@20 & R@20 & N@20 & R@20 & N@20\\
\midrule
\midrule[0.1pt]
  \multirow{2}*{\textsl{LB}}    &{24.24} & {43.15}   & {17.68}& {24.23}  & {15.31}& {15.44}  & {6.24} & {10.98} & {4.83} & {7.98}\\
  ~        &\textit{\drop {-8.60\%}}  &\textit{\drop {-7.84\%}}  &\textit{\drop {-5.25\%}}  &\textit{\drop {-3.96\%}}  &\textit{\drop {-6.65\%}}  &\textit{\drop {-11.42\%}}  &\textit{\drop {-8.24\%}}  &\textit{\drop {-8.65\%}} &\textit{\drop {-6.40\%}}  &\textit{\drop  {-9.11\%}}\\
  \midrule[0.1pt]
{\textbf{\model} }  &\textbf{26.52}& \textbf{46.82}   & \textbf{18.66}& \textbf{25.23}  & \textbf{16.40}& \textbf{17.43}  & \textbf{6.80}& \textbf{12.02} & \textbf{5.16}& \textbf{8.78}\\
\bottomrule
\end{tabular}}
\end{table}

\subsubsection{\textbf{Implementation of $w_l$.}}
\label{app:wl}
We try three additional implementations of $w_l$ and report the results in Tables~\ref{tab:wl}.
\begin{enumerate}[leftmargin=*]
\item {$w_l$ = $\frac{1}{L+1}$} equally contributes for all embedding segments.

\item {$w_l$ = $\frac{1}{L+1-l}$} is positively correlated to $l$, so as to highlight higher-order structures of the interaction graph. 

\item {$w_l$ = $2^{-(L+1-l)}$} is positively correlated to $l$ with exponentiation.
\end{enumerate}
The experimental results indicate that Implementation (2) performs relatively well compared to the others, underscoring the significance of leveraging higher-order graph information. This supports the design choice in \model, where {$w_l$ $\propto$ $l$}. Despite its simplicity, this approach proves to be both effective and yields superior recommendation accuracy.

\begin{table}[t]
\centering
\caption{\new Implementation of $w_l$.}
\label{tab:wl}
{\new
\begin{tabular}{c | c c | c c | c c | c c | c c }
\toprule
\multirow{2}*{Implementation}  & \multicolumn{2}{c|}{MovieLens} & \multicolumn{2}{c|}{Gowalla} & \multicolumn{2}{c|}{Pinterest} & \multicolumn{2}{c|}{Yelp2018} & \multicolumn{2}{c}{Amazon-Book}\\
               ~  & R@20 & N@20 & R@20 & N@20 & R@20 & N@20 & R@20 & N@20 &R@20 & N@20\\
\midrule[0.1pt]
\midrule[0.1pt]
  (1)           &{23.83}  &{42.24} &{16.83} &{22.95} &{14.68} &{16.04} &{5.83} &{10.44} &{4.73} &{7.72} \\
  (2)           &{25.72}  &{45.13} &{17.92} &{24.83} &{15.74} &{16.93} &{6.62} &{11.57} &{4.69} &{7.93} \\        
  (3)           &{22.67}  &{39.28} &{15.80} &{21.83} &{12.99} &{15.33} &{5.78} &{9.67} &{3.83} &{7.16} \\  
\midrule[0.1pt]
\textbf{\model}     &\textbf{26.52}& \textbf{46.82}   & \textbf{18.66}& \textbf{25.23}  & \textbf{16.40}& \textbf{17.43}  & \textbf{6.80}& \textbf{12.02} & \textbf{5.16}& \textbf{8.78}\\
\bottomrule
\end{tabular}}
\end{table}

\subsubsection{\textbf{Implementation of $w_k$.}}
\label{app:wk}
We further evaluate different $w_k$:
\begin{enumerate}[leftmargin=*]
\item  {$w_k$ = $\frac{R-k}{R}$} is negatively correlated to the ranking position $k$.
\item  {$w_k$ = $\frac{1}{k}$} is inversely proportional to position $k$.
\item  {$w_k$ = $2^{-k}$} is exponential to the value of $-k$.
\end{enumerate}

As shown in Table~\ref{tab:wk}, Implementation (3) performs slightly worse than Equation~(\ref{eq:wk}) but generally outperforms the other two methods. 
This highlights the effectiveness of exponential modeling in capturing the contribution of items for approximating tailed item popularity~\cite{rendle2014improving}. Additionally, Equation~(\ref{eq:wk}) introduces hyper-parameters, offering the flexibility to adjust the function's properties across different datasets.

\begin{table}[t]
\centering
\caption{\new Implementation of $w_k$.}
\label{tab:wk}
{\new 
\begin{tabular}{c | c c | c c | c c | c c | c c }
\toprule
 \multirow{2}*{Implementation} & \multicolumn{2}{c|}{MovieLens} & \multicolumn{2}{c|}{Gowalla} & \multicolumn{2}{c|}{Pinterest} & \multicolumn{2}{c|}{Yelp2018} & \multicolumn{2}{c}{Amazon-Book}\\
               ~  & R@20 & N@20 & R@20 & N@20 & R@20 & N@20 & R@20 & N@20 &R@20 & N@20\\
\midrule[0.1pt]
\midrule[0.1pt]
  (1)           &{24.83}  &{44.80} &{18.11} &{25.13} &{16.13} &{16.89} &{6.35} &{11.46} &{4.73} &{7.91} \\
  (2)           &{25.38}  &{45.34} &{18.13} &{25.09} &{15.84} &{16.45} &{6.28} &{11.42} &{4.77} &{8.34} \\        
  (3)           &{25.24}  &{45.23} &{18.44} &{24.97} &{16.14} &{16.94} &{6.54} &{11.50} &{4.81} &{8.46} \\  
\midrule[0.1pt]
\textbf{\model}    &\textbf{26.52}& \textbf{46.82}   & \textbf{18.66}& \textbf{25.23}  & \textbf{16.40}& \textbf{17.43}  & \textbf{6.80}& \textbf{12.02} & \textbf{5.16}& \textbf{8.78}\\
\bottomrule
\end{tabular}}
\end{table}

\section{Related Work}
\label{sec:work}

\paragraph{\textbf{Full-precision Recommender Models.}} (1) \textit{Collaborative Filtering (CF)} is a widely used approach in modern recommender systems~\cite{covington2016deep,pinsage,yang2022hrcf,lin2024effective,zhang2022knowledge,luo2025rank}. Earlier CF methods, such as \textit{Matrix Factorization}~\cite{koren2009matrix,rendle2012bpr}, focus on reconstructing historical interactions to learn user-item embeddings. More recent models, like NeurCF~\cite{neurcf} and attention-based approaches~\cite{chen2017attentive,he2018nais}, improve performance by leveraging neural networks. (2) \textit{Graph-based} methods have been widely used in many domains~\cite{wu2023survey,zhang2024geometric,zhang2023contrastive}. In recommendation, they explore the interaction graph structure for knowledge learning. 
Graph convolutional networks (GCNs)~\cite{graphsage,kipf2016semi} propagate knowledge via graph topologies~\cite{wang2022aep,chen2025semi} and have inspired both early methods, such as GC-MC~\cite{berg2017graph} and PinSage~\cite{pinsage}, as well as recent models like NGCF~\cite{ngcf}, DGCF~\cite{dgcf}, and LightGCN~\cite{lightgcn}, which effectively capture higher-order collaborative filtering signals among high-hop neighbors for improved recommendations.

\paragraph{\textbf{Learning to Hash.}} Hashing-based methods convert dense floating-point embeddings into binary spaces to accelerate \textit{Approximate Nearest Neighbor} (ANN) searches. A prominent model, LSH~\cite{lsh}, has inspired numerous approaches across various domains, including fast image retrieval~\cite{hashnet}, document search~\cite{li2014two}, and categorical information retrieval~\cite{kang2021learning}. In Top-K recommendation, early models~\cite{zhang2016discrete,zhang2017discrete,li2019learning} incorporate neural network architectures. CIGAR~\cite{kang2019candidate} further refines these methods with adaptive designs for fast candidate generation. HashGNN~\cite{hashgnn} integrates hashing techniques with graph neural networks~\cite{wu2020comprehensive,wang2024uncertainty} to capture the graph structure in user-item interactions for recommendation. However, relying solely on binary codes often leads to significant performance degradation. To address this, CIGAR includes additional full-precision recommender models (e.g., BPR-MF~\cite{rendle2012bpr}) for fine-grained re-ranking, while HashGNN introduces a relaxed version that mixes full-precision and binary embedding codes.

\paragraph{\textbf{Quantization-based Models.}} Quantization-based models share techniques with hashing-based methods, often using $\sign(\cdot)$ due to its simplicity. Unlike hashing-based methods, however, quantization models are not focused on extreme compression, instead utilizing multi-bit, 2-bit, and 1-bit quantization to optimize performance~\cite{qiu2024hihpq,chen2021towards}. Recently, attention has shifted toward quantizing graph-based models, such as Bi-GCN~\cite{bigcn} and BGCN~\cite{bahri2021binary}. However, these models are primarily designed for geometric classification tasks, leaving their effectiveness in product recommendation unclear. 
In response, we introduce BiGeaR~\cite{chen2022learning}, a model that learns 1-bit user-item representation quantization for Top-K recommendation. 
Building on this, we propose \model~with enhanced design features aimed at further improving both efficiency and model performance.

{\rev \paragraph{\textbf{Knowledge Distillation}.} 
Knowledge Distillation (KD) represents the process of transferring knowledge from a larger model to a smaller one~\cite{hinton2015distilling}.
We review KD techniques specifically for the topics of Graph Neural Networks (GNN) and Learning to hash as follows.
(1) KD in GNNs focuses on transferring knowledge from a large, complex model (teacher) to a smaller, more efficient model (student) while maintaining performance~\cite{tian2023knowledge,yang2020distilling,deng2021graph}.
G-CRD~\cite{joshi2022representation} leverages contrastive learning methods~\cite{wang2024graph,zhang2023contrastive} to better capture topological information by aligning teacher and student node embeddings in a shared representation space.
HKD~\cite{zhou2021distilling} uses GNNs to integrate both individual knowledge and relational knowledge, i.e., two types of knowledge, while reserving their inherent correlation in the distillation process.
KDGA~\cite{wu2022knowledge} focuses on addressing the negative augmentation problem in graph structure augmentation.
A recent work~\cite{liu2024fine} further optimizes the resource consumption of GNNs by proposing a KD method for fine-grained learning behavior.
FreeKD~\cite{feng2022freekd} considers reinforcement learning in KD for GNNs.
(2) KD is also widely used in learning to hash to reduce the information discrepancy and balance efficiency.
For example, UKD~\cite{hu2020creating} and SKDCH~\cite{su2021semi} apply KD in cross-modal hashing to reduce modality discrepancy.
\citet{jang2022deep} introduce a self-distilled hashing scheme with data augmentation designs.
HMAH~\cite{tan2022teacher} constructs a hierarchical message aggregation mechanism to better align the heterogeneous modalities and model the fine-grained multi-modal correlations. 
A recent work~\cite{yu2024unsupervised} introduces KD to improve the effectiveness of large-scale cross-media hash retrieval. 
Generally, Combining KD with learning to hash allows the student model to benefit from the teacher's superior representation capabilities while maintaining the efficiency of compact representations.
}

{\new
\section{Conclusions}
\label{sec:con}
In this paper, we present \model, a framework for learning binarized graph representations for recommendation through a series of advanced binarization techniques. 
Building upon its predecessor, \model~specifically leverages the supervisory signals from pseudo-positive item samples and synthesized embeddings for learning enrichment. 
Extensive experiments validate \model's performance superiority, demonstrating that the new design elements integrate seamlessly with other components. 
Comprehensive empirical analyses further confirm the effectiveness of \model~compared to other binarization-based recommender systems, achieving state-of-the-art performance.
As for future work, we consider to leverage Large Language Models (LLMs)~\cite{qiu2024ease,wei2024llmrec} and multimodal information~\cite{li2024benchmarking,liang2024survey} in recommedner systems for optimization.
}



\bibliographystyle{ACM-Reference-Format}
{
\bibliography{ref}
}

\end{document}